\documentclass[11pt,a4paper]{article}
\usepackage{jheppub}

\usepackage[sort&compress]{natbib}
\usepackage{graphicx}
\usepackage{dcolumn}
\usepackage{bm}

\usepackage{subfigure}

\usepackage{amsmath}
\usepackage{amssymb}
\usepackage{amscd}
\usepackage{epsfig}
\usepackage{latexsym,amssymb}

\usepackage{dcolumn}
\usepackage{bm}

\newcommand{\ttbar}{\ensuremath{t\bar{t}}}
\newcommand{\bbar}{\ensuremath{b\bar{b}}}
\newcommand{\bbbar}{\ensuremath{b\bar{b}}}
\newcommand{\ppbar}{\ensuremath{p\bar{p}}}
\newcommand{\qqbar}{\ensuremath{q\bar{q}}}

\newcommand{\invfb}{ \ensuremath{\rm fb^{-1}}}

\newcommand{\gevc}{\ensuremath{ \text{GeV/}c}}
\newcommand{\gevcc}{\ensuremath{ \text{GeV/}c^{\text{2}}}}

\newcommand{\mbb}{\ensuremath{m_{bb}}}
\newcommand{\mqq}{\ensuremath{m_{qq}}}

\newcommand{\et}{\ensuremath{E_T}}
\newcommand{\pt}{\ensuremath{p_T}}

\newcommand{\Tag}{{TAG}}
\newcommand{\Ctrl}{{CONTROL}}
\newcommand{\Njet}{{NJET6}}

\newcommand{\SD}{Higgs-NN}

\def\fourJet15{\texttt{L2\_FOUR\_JET15} }
\def\SUMET125{\texttt{L2\_SUMET125} }
\def\SUMET175{\texttt{L2\_SUMET175} }

\preprint{CDF/DOC/EXOTIC/CDFR/10891}

\title{Search for the Higgs boson in the all-hadronic final state using the full CDF data set}

\collaboration{CDF Collaboration}

\affiliation[1]{Institute of Physics, Academia Sinica, Taipei, Taiwan 11529, Republic of China}
\affiliation[2]{Argonne National Laboratory, Argonne, Illinois 60439, USA}
\affiliation[3]{University of Athens, 157 71 Athens, Greece}
\affiliation[4]{Institut de Fisica d'Altes Energies, ICREA, Universitat Autonoma de Barcelona, E-08193, Bellaterra (Barcelona), Spain}
\affiliation[5]{Baylor University, Waco, Texas 76798, USA}
\affiliation[6]{Istituto Nazionale di Fisica Nucleare Bologna, Italy}
\affiliation[7]{University of California, Davis, Davis, California 95616, USA}
\affiliation[8]{University of California, Los Angeles, Los Angeles, California 90024, USA}
\affiliation[9]{Instituto de Fisica de Cantabria, CSIC-University of Cantabria, 39005 Santander, Spain}
\affiliation[10]{Carnegie Mellon University, Pittsburgh, Pennsylvania 15213, USA}
\affiliation[11]{Enrico Fermi Institute, University of Chicago, Chicago, Illinois 60637, USA}
\affiliation[12]{Comenius University, 842 48 Bratislava, Slovakia; Institute of Experimental Physics, 040 01 Kosice, Slovakia}
\affiliation[13]{Joint Institute for Nuclear Research, RU-141980 Dubna, Russia}
\affiliation[14]{Duke University, Durham, North Carolina 27708, USA}
\affiliation[15]{Fermi National Accelerator Laboratory, Batavia, Illinois 60510, USA}
\affiliation[16]{University of Florida, Gainesville, Florida 32611, USA}
\affiliation[17]{Laboratori Nazionali di Frascati, Istituto Nazionale di Fisica Nucleare, I-00044 Frascati, Italy}
\affiliation[18]{University of Geneva, CH-1211 Geneva 4, Switzerland}
\affiliation[19]{Glasgow University, Glasgow G12 8QQ, United Kingdom}
\affiliation[20]{Harvard University, Cambridge, Massachusetts 02138, USA}
\affiliation[21]{Division of High Energy Physics, Department of Physics, University of Helsinki and Helsinki Institute of Physics, FIN-00014, Helsinki, Finland}
\affiliation[22]{University of Illinois, Urbana, Illinois 61801, USA}
\affiliation[23]{The Johns Hopkins University, Baltimore, Maryland 21218, USA}
\affiliation[24]{Institut f\"{u}r Experimentelle Kernphysik, Karlsruhe Institute of Technology, D-76131 Karlsruhe, Germany}
\affiliation[25]{Center for High Energy Physics: Kyungpook National University, Daegu 702-701, Korea; Seoul National University, Seoul 151-742, Korea; Sungkyunkwan University, Suwon 440-746, Korea; Korea Institute of Science and Technology Information, Daejeon 305-806, Korea; Chonnam National University, Gwangju 500-757, Korea; Chonbuk National University, Jeonju 561-756, Korea}
\affiliation[26]{Ernest Orlando Lawrence Berkeley National Laboratory, Berkeley, California 94720, USA}
\affiliation[27]{University of Liverpool, Liverpool L69 7ZE, United Kingdom}
\affiliation[28]{University College London, London WC1E 6BT, United Kingdom}
\affiliation[29]{Centro de Investigaciones Energeticas Medioambientales y Tecnologicas, E-28040 Madrid, Spain}
\affiliation[30]{Massachusetts Institute of Technology, Cambridge, Massachusetts 02139, USA}
\affiliation[31]{Institute of Particle Physics: McGill University, Montr\'{e}al, Qu\'{e}bec, Canada H3A~2T8; Simon Fraser University, Burnaby, British Columbia, Canada V5A~1S6; University of Toronto, Toronto, Ontario, Canada M5S~1A7; and TRIUMF, Vancouver, British Columbia, Canada V6T~2A3}
\affiliation[32]{University of Michigan, Ann Arbor, Michigan 48109, USA}
\affiliation[33]{Michigan State University, East Lansing, Michigan 48824, USA}
\affiliation[34]{Institution for Theoretical and Experimental Physics, ITEP, Moscow 117259, Russia}
\affiliation[35]{University of New Mexico, Albuquerque, New Mexico 87131, USA}
\affiliation[36]{The Ohio State University, Columbus, Ohio 43210, USA}
\affiliation[37]{Okayama University, Okayama 700-8530, Japan}
\affiliation[38]{Osaka City University, Osaka 588, Japan}
\affiliation[39]{University of Oxford, Oxford OX1 3RH, United Kingdom}
\affiliation[40]{Istituto Nazionale di Fisica Nucleare, Sezione di Padova-Trento, Italy}
\affiliation[41]{University of Pennsylvania, Philadelphia, Pennsylvania 19104, USA}
\affiliation[42]{Istituto Nazionale di Fisica Nucleare Pisa, Italy}
\affiliation[43]{University of Pittsburgh, Pittsburgh, Pennsylvania 15260, USA}
\affiliation[44]{Purdue University, West Lafayette, Indiana 47907, USA}
\affiliation[45]{University of Rochester, Rochester, New York 14627, USA}
\affiliation[46]{The Rockefeller University, New York, New York 10065, USA}
\affiliation[47]{Istituto Nazionale di Fisica Nucleare, Sezione di Roma 1, Italy}
\affiliation[48]{Rutgers University, Piscataway, New Jersey 08855, USA}
\affiliation[49]{Texas A\&M University, College Station, Texas 77843, USA}
\affiliation[50]{Istituto Nazionale di Fisica Nucleare Trieste/Udine, I-34100 Trieste, Italy}
\affiliation[51]{University of Tsukuba, Tsukuba, Ibaraki 305, Japan}
\affiliation[52]{Tufts University, Medford, Massachusetts 02155, USA}
\affiliation[53]{University of Virginia, Charlottesville, Virginia 22906, USA}
\affiliation[54]{Waseda University, Tokyo 169, Japan}
\affiliation[55]{Wayne State University, Detroit, Michigan 48201, USA}
\affiliation[56]{University of Wisconsin, Madison, Wisconsin 53706, USA}
\affiliation[57]{Yale University, New Haven, Connecticut 06520, USA}

\affiliation[]{\vspace{1cm} With visitors from}
\affiliation[a]{Istituto Nazionale di Fisica Nucleare, Sezione di Cagliari, 09042 Monserrato (Cagliari), Italy}
\affiliation[b]{University of CA Irvine, Irvine, CA 92697, USA}
\affiliation[c]{University of CA Santa Barbara, Santa Barbara, CA 93106, USA}
\affiliation[d]{University of CA Santa Cruz, Santa Cruz, CA 95064, USA}
\affiliation[e]{Institute of Physics, Academy of Sciences of the Czech Republic, Czech Republic}
\affiliation[f]{CERN, CH-1211 Geneva, Switzerland}
\affiliation[g]{Cornell University, Ithaca, NY 14853, USA}
\affiliation[h]{University of Cyprus, Nicosia CY-1678, Cyprus}
\affiliation[i]{Office of Science, U.S. Department of Energy, Washington, DC 20585, USA}
\affiliation[j]{University College Dublin, Dublin 4, Ireland}
\affiliation[k]{ETH, 8092 Zurich, Switzerland}
\affiliation[l]{University of Fukui, Fukui City, Fukui Prefecture, Japan 910-0017}
\affiliation[m]{Universidad Iberoamericana, Mexico D.F., Mexico}
\affiliation[n]{University of Iowa, Iowa City, IA 52242, USA}
\affiliation[o]{Kinki University, Higashi-Osaka City, Japan 577-8502}
\affiliation[p]{Kansas State University, Manhattan, KS 66506, USA}
\affiliation[q]{Ewha Womans University, Seoul, 120-750, Korea}
\affiliation[r]{University of Manchester, Manchester M13 9PL, United Kingdom}
\affiliation[s]{Queen Mary, University of London, London, E1 4NS, United Kingdom}
\affiliation[t]{University of Melbourne, Victoria 3010, Australia}
\affiliation[u]{Muons, Inc., Batavia, IL 60510, USA}
\affiliation[v]{Nagasaki Institute of Applied Science, Nagasaki, Japan}
\affiliation[w]{National Research Nuclear University, Moscow, Russia}
\affiliation[x]{Northwestern University, Evanston, IL 60208, USA}
\affiliation[y]{University of Notre Dame, Notre Dame, IN 46556, USA}
\affiliation[z]{Universidad de Oviedo, E-33007 Oviedo, Spain}
\affiliation[aa]{CNRS-IN2P3, Paris, F-75205 France}
\affiliation[bb]{Texas Tech University, Lubbock, TX 79609, USA}
\affiliation[cc]{Universidad Tecnica Federico Santa Maria, 110v Valparaiso, Chile}
\affiliation[dd]{Yarmouk University, Irbid 211-63, Jordan}
\affiliation[ee]{University of Bologna, I-40127 Bologna, Italy}
\affiliation[ff]{University of Padova, I-35131 Padova, Italy}
\affiliation[gg]{University of Pisa} 
\affiliation[hh]{University of Siena}
\affiliation[ii]{Scuola Normale Superiore, I-56127 Pisa, Italy}
\affiliation[jj]{California Institute of Technology, Pasadena, California 91125, USA}
\affiliation[kk]{INFN Pavia and University of Pavia, Italy}

\author[21]{T.~Aaltonen}
\author[9]{B.~\'{A}lvarez~Gonz\'{a}lez}
\author[40]{S.~Amerio}
\author[32]{D.~Amidei}
\author[15]{A.~Anastassov}
\author[17]{A.~Annovi}
\author[12]{J.~Antos}
\author[15]{G.~Apollinari}
\author[15]{J.A.~Appel}
\author[44,jj]{A.~Apresyan}
% ...

\date{\today}

\abstract{

This paper reports the result of a search for the standard model Higgs 
boson in events containing four reconstructed jets associated with quarks.
For masses below 135\,\gevcc, the Higgs boson decays to bottom-antibottom quark pairs
are dominant and result primarily in two hadronic jets.  An additional two jets can be produced in the
hadronic decay of a $W$ or $Z$ boson produced in association with the Higgs boson, 
or from the incoming quarks that produced the Higgs boson through the vector-boson fusion process.
The search is performed using a sample of $\sqrt{s}= 1.96$ TeV proton-antiproton collisions 	
corresponding to an integrated luminosity of
9.45\,\invfb\, recorded by
the CDF II detector.
The data are in agreement with the background model and 
95\% credibility level upper limits on Higgs boson production are set as a function of the
Higgs boson mass. 
The median expected (observed) limit 
for a 125\,\gevcc\,\,Higgs boson is
11.0 (9.0) times the predicted standard model rate.
}

\keywords{Higgs, All-Hadronic, $b$-jets}

\begin{document}
\maketitle
\flushbottom

%
\section{Introduction}
The Higgs boson is the physical manifestation of the hypothesized
mechanism that provides mass to fundamental
particles in the standard model (SM)~\cite{Higgs:1964ia,Englert1964Broken-Symmetry,PhysRevLett.13.508}.
Direct searches at the LEP collider~\cite{Barate:2003sz}, the
Tevatron~\cite{Tevatron:Higgs}, and the
LHC~\cite{CMSHiggsObservation,ATLASHiggsObservation} have excluded SM
Higgs boson masses at the 95\% confidence level or 95\%
credibility level (CL), except within the range 122-128\,\gevcc.
The most sensitive searches at the LHC are based on SM Higgs boson decays to pairs of gauge bosons. 
At the Tevatron,  searches based on Higgs boson decay to bottom-antibottom quark pairs (\bbbar)
are the most sensitive within the allowed range.
Searches in  this channel offers complementary information on fermion Yukawa couplings to the Higgs boson.

Recently, the ATLAS and CMS collaborations have reported the observation of a Higgs-like particle 
at a mass of $\approx125$\,\gevcc~\cite{CMSHiggsObservation,ATLASHiggsObservation}, and 
the Tevatron has reported evidence for a particle decaying to \bbbar\,  
produced in association with a $W/Z$ boson for masses within the range
120 - 135\,\gevcc~\cite{Aaltonen:2012qt}. 

This paper describes a search for the Higgs boson
using a data sample corresponding to an integrated luminosity of
9.45\invfb\, of \ppbar\,\,collisions  at $\sqrt{s} = 1.96$\,TeV recorded by the Collider Detector at Fermilab (CDF II).
In this search two production mechanisms are studied: associated
vector-boson 
production (\!{\it VH}) and vector-boson fusion (VBF).
The {\it VH} channel denotes the process 
$\ppbar \rightarrow W/Z + H \rightarrow q\bar{q}^{\prime} + \bbbar$. 
The VBF channel identifies the process 
$\ppbar \rightarrow q\bar{q}^{\prime} H \rightarrow q\bar{q}^{\prime} \bbbar$, 
where the two incoming quarks each radiate a weak boson,
which subsequently fuse into a Higgs boson.
In both channels, the Higgs boson decays to \bbbar,
and is produced in association with two other quarks ($q\bar{q}^{\prime}$).
Data are tested against the hypothesis of the presence of Higgs boson with mass 
in the range $100 < m_H < 150\,\gevcc$. 
The $H \rightarrow \bbbar$ mode is the dominant decay for 
$m_H < 135\,\gevcc$~\cite{Djouadi:1997yw}.


Searches for a Higgs boson performed in final states containing leptons,
jets, and missing energy have the advantage of
smaller background, but the
Higgs boson signal yields are also very small.
The  all-hadronic search channel, described here, has larger potential signal contributions
but suffers from substantial QCD multi-jet background contributions.
The challenge of this channel is to accurately model and reduce 
 the multi-jet background. 
Two previous papers were published on searches for a Higgs boson in
the all-hadronic channel at CDF using data sets of
2\invfb\,~\cite{PhysRevLett.103.221801} and
4\invfb\,~\cite{PhysRevD.84.052010}. 
Another paper was published on searches for a Higgs boson in
the all-hadronic channel at CDF using data collected during Run I~\cite{AllHadHiggsRunI}.
The LEP collider also conducted searches 
for the Higgs boson in the all-hadronic final state 
in the
 $e^{+}e^{-} \rightarrow ZH \rightarrow \qqbar + \bbbar$ channel~\cite{Barate:2003sz}.

%
\section{The CDF II detector}
The CDF II detector is an azimuthally and forward-backward symmetric
multipurpose detector.
CDF II uses a cylindrical coordinate 
system with the $z$-axis aligned along the proton beam 
direction, where $\theta$ is the polar angle relative to the $z$-axis and $\phi$ 
is the azimuthal angle relative to the $x$-axis. The pseudorapidity 
is defined as $\eta = -\ln(\tan \theta /2$) and the transverse 
energy is calculated as $E_T = E \sin \theta$.

The CDF II detector consists of  a pair of concentric charged-particle tracking detectors immersed 
in a 1.4\,T solenoid magnetic field, surrounded by calorimeters and muon detectors. 
The inner tracking detector is the silicon vertex detector that is located immediately outside the beam pipe,
provides precise three-dimensional reconstruction of charged-particle trajectories (tracks)
and is used
to identify displaced vertices associated with bottom-quark and charm-quark
hadron decays. 
%
%
%
%
The momenta of charged particles is measured precisely in the central
outer tracker (COT), a cylindrical multiwire drift chamber.
The tracking detectors cover the pseudorapidity range $|\eta| < 1.1$. 
Outside the COT are electromagnetic and hadronic calorimeters arranged
in a projective-tower geometry, covering the region $|\eta| < 3.5$, to provide energy measurements for both charged and neutral particles. 
Drift chambers and scintillator counters in the region $|\eta| < 1.5$
provide muon identification outside the calorimeters. 
Luminosity is measured using low-mass gaseous Cherenkov luminosity counters
(CLC). There are two CLC modules in the CDF II detector installed at
small angles in the proton and antiproton directions, arranged in three concentric layers around the beam pipe.
More details about the CDF II detector can be found in refs.~\cite{Acosta:2004yw,Acosta:2004hw,Abulencia:2005ix}.

Jets are defined by a cluster of energy deposited in the calorimeter using a jet clustering
algorithm (JetClu)~\cite{Abe:1991ui} with a cone of fixed radius.
The JetClu algorithm begins by creating a list of calorimeter towers above a
fixed $E_T$ threshold to be used as seeds for the jet finder.
This threshold is set to $1.0$ GeV. 
Preclusters are formed from an unbroken chain of contiguous seed
towers with a continuously decreasing tower $E_T$. If a tower is
outside a  window of seven towers surrounding the seed,  it is  used  to form a  new
precluster. These preclusters are used as a starting point for cone
clustering. 
First, the $E_T$ weighted centroid of the
precluster is found and a  cone in $\eta - \phi$ space
of radius $R$ is formed around the centroid. For this analysis,
$\Delta R = \sqrt{(\Delta \phi)^2 + (\Delta \eta)^2} = 0.4$. Then, all towers with 
an $E_T$ of, at least, $100$ MeV are incorporated into the cluster.  A tower is included in a cluster if 
its centroid is inside the cone, otherwise it is excluded.  A new cluster center is calculated 
from the set of towers within the clustering cone,  again using an $E_T$ weighted centroid, and 
a new cone is drawn about this position. The process of recomputing a centroid and finding 
new or deleting old towers is iterated until the tower list remains unchanged.
Corrections are applied to the measured jet energy to account for
detector calibrations, multiple interactions, underlying event, and
energy outside of the jet cone~\cite{Bhatti:2005ai}.

The data for this analysis are collected using two online event selections (triggers). Events in the first
3.0\invfb\, are triggered by selecting those containing
at least four jets
with $E_T \geq 15$ GeV and 
total calorimeter $E_T$ greater than $175$ GeV. Events in the
remaining 6.45\invfb\,  are selected by requiring 
at least three jets
with $E_T \geq 20$ GeV and total calorimeter $E_T$ greater than $130$ GeV.

%
\section{Event selection}~\label{SECTION:EventSelection}
Events with isolated leptons or missing transverse energy
significance\footnote{Missing transverse energy significance is
  defined as the ratio of the missing transverse energy to the
  square root of the total transverse energy.
    The missing transverse energy, $\not\!\!E_T=|{\not\!\!\vec{E}_T}|$,
    where $\not\!\!\vec{E}_T$ is defined by, 
$\not\!\!\vec{E}_T = - \sum_{i} E_T^i \hat{n}_i$, where
$i$ is calorimeter tower number with $|\eta| < 3.6$, 
$\hat{n}_i$ is a unit vector perpendicular to the beam axis and pointing at the i$^{th}$ calorimeter tower.
  } 
  greater than 6.0, which is indicative of the presence of neutrinos,
     are removed to
ensure an event sample
independent from 
other Higgs boson searches at CDF.
Events containing four or five jets,  
with $E_{T}>15$\,GeV and
$|\eta|< 2.4$ are selected. 

To reduce the QCD multi-jet background, exactly two bottom-quark jets
($b$ jets) are required.
Two algorithms are
used to identify $b$ jets: the SecVtx algorithm~\cite{Acosta:2004hw} and the JetProb
algorithm~\cite{Abulencia:2006kv}.
The SecVtx algorithm attempts
to reconstruct 
the secondary vertex 
associated with a 
bottom-quark ($b$) decay. The JetProb
algorithm searches the impact parameter of
the charged-particle trajectories (tracks) within a jet and
selects those that are inconsistent with originating from the decay of
a particle occurred in the vicinity of the primary event vertex. 
An additional energy correction is applied to jets identified as
$b$ jets (section~\ref{SECTION:bJetEnergyCorrection}). 
Untagged jets (non~$b$ jets) are referred to as $q$ jets in this paper.

All selected jets are ordered in
 \et\, and the four highest \et\, jets are considered. 
The scalar sum of the four selected jet \et\,\!s
 (SumEt) is required to exceed 220 GeV and two of the four selected jets must be $b$ jets.



The signal-to-background ratio is enhanced by dividing the data into two
independent $b$-tagging
categories: SS in which both jets are tagged by SecVtx, and SJ in which
one jet is tagged by SecVtx and the other by JetProb. 
If a jet is tagged by both algorithms, it is classified as tagged by
SecVtx because of the lower misidentification rate.
Events in which both jets are tagged only by JetProb are not used because the 
increase in background contributions is substantially larger than that for the signal.

The signal region is defined by requirements on the invariant mass of the two $b$-tagged jets
(\mbb) and the two untagged jets (\mqq).
The {\it VH} channel features two intermediate resonances, one from
the potential Higgs boson decay,
 in \mbb, and another from the $W/Z$ decay,  in \mqq. The
VBF channel shares the same \mbb\, resonance but the two $q$ jets are
not produced from the decay of a particle. However these two $q$ jets tend to be produced
with large $\eta$ separation which gives a large effective \mqq\, mass. 
The Higgs boson search region is defined as $75<\mbb<175\,\gevcc$ and
$\mqq>50\,\gevcc$.




%
\section{Signal and background samples}
Backgrounds that contribute to the $qq\bbbar$ final state originate from
QCD multi-jet production, top-quark pair production, single-top-quark production, 
$W \rightarrow q'\bar{q}$ plus $\bbar$ or charm-quark pair
($c\bar{c}$) production ($W+$$\mathit{HF}$), 
$Z \rightarrow \bbar$,$c\bar{c}$ plus jets production ($Z$+jets),
and diboson production ($WW$, $WZ$, $ZZ$).
About 98\% of the total background comes from QCD multi-jet production. 
Signal and non-QCD backgrounds  yields are estimated from Monte Carlo
(MC) simulation.
The $W+$$\mathit{HF}$ and $Z+$jets contributions are modeled with the {\sc  alpgen}~\cite{Mangano:2002ea} generator
for simulating the bosons
plus parton production, and  {\sc pythia}~\cite{Sjostrand:2000wi} for modeling parton  showers.  
The other non-QCD backgrounds and the signal are modeled with  {\sc  pythia}~\cite{Sjostrand:2000wi}.
%
%
%
%
All MC-simulated samples use the
{\sc CTEQ5L}~\cite{CTEQ5L} parton
distribution function (PDF) at leading order (LO) 
and 
are processed through the full CDF
II detector simulation~\cite{2003physics...6031G}, based on {\sc geant}~\cite{GEANT}, 
that includes the trigger simulation and their trigger efficiencies are corrected 
as described in ref.~\cite{PhysRevLett.103.221801}.




The expected signal yield in the SS (SJ) channel is $27.1\pm4.1$ ($9.1\pm1.4$)
for $m_H = 125\,\gevcc$.
The selected number of data events for SS (SJ) are 87272 (46818).
A data-driven model is
used to predict the shape of QCD multi-jet background but not the 
overall yield (section~\ref{SECTION:QCDMODEL}). 
The number of QCD multi-jet events in each channel
is estimated as the difference
between the number of 
data events and the predicted number of non-QCD events
estimated with MC (neglecting the potential Higgs boson contribution).
Expected and observed event yields are summarized in table~\ref{TABLE:NumberOfBackgroundEvents}.
In the final fit used to extract a potential Higgs boson signal, the overall normalization of the QCD multi-jet
background is treated as an unconstrained parameter.



%
%


\begin{table}[h!]
\begin{center}
\begin{tabular}{|l|c|c|} 
\hline
Backgrounds               &             SS channel        &            SJ channel       \\
\hline
$t\bar{t}$                    &     $1032\pm156$   &     $384\pm 57$  \\
Single top $s$ channel                                  &      $111\pm19$    &     $ 38\pm6$    \\
Single top $t$ channel                                  &      $ 44\pm7$     &      $26\pm4$    \\
$W+b\bar{b}$                                             &       $77\pm40$    &      $29\pm15$   \\
$W+c\bar{c}$                                              &        $8\pm4$     &       $7\pm4$    \\
$Z(\rightarrow b\bar{b} / c\bar{c})$+jets    &      $873\pm452$   &     $338\pm175$  \\
$WW$                                                              &        $6\pm1$     &       $6\pm1$   \\
$WZ$                                                               &       $20\pm3$     &       $8\pm1$    \\
$ZZ$                                                               &       $21\pm3$     &       $8\pm1$    \\ \hline
Total non-QCD                                           &     $2192\pm480$   &     $844\pm185$  \\ \hline \hline
Data                                                           &     87272              &    46818             \\ \hline
QCD multi-jet                                                          &     85080            &    45974           \\ \hline
\hline
Higgs signal (125\,\gevcc)       &     $27\pm4$       &     $9\pm1$     \\
\hline
\end{tabular}
\end{center}
\caption{Expected number of background and signal ($m_H = 125\,\gevcc$) events that pass the complete event
selection for the SS and SJ $b$-tag categories. 
The number of QCD multi-jet events is
estimated as the difference between data and predicted non-QCD
backgrounds (neglecting the potential Higgs contribution). 
The uncertainties of the signal and non-QCD background rate predictions include statistical
and systematic rate uncertainties, such as cross-section and integrated luminosity, 
as described in section~\ref{SEC:Systematics}.
%
}
\label{TABLE:NumberOfBackgroundEvents}
\end{table}

%
\section{Search strategy}

The main challenge is to 
accurately model and reduce the QCD multi-jet background. 
The modeling of this background   
is obtained from a data-driven 
technique described in section~\ref{SECTION:QCDMODEL}.
This avoids the need of generating large volumes of QCD multi-jet simulation samples, 
which is computationally intensive 
and unlikely to accurately reproduce the multi-jet spectrum.


The overwhelming QCD multi-jet background is suppressed by relying on multi-variate techniques
that combine information from multiple variables to identify potential Higgs boson events.  
For example, the best signal-to-background ratio using just \mbb\, is 0.0007\footnote{Only considering events under the Higgs signal peak ($120 < \mbb < 140\,\gevcc$), figure~\ref{FIG:TrainingVariables_SS_1}\subref{mass_bb}.}.  
In this search, the use of
multi-variate techniques improves the  signal-to-background to 0.006\footnote{Only considering events with Higgs-NN $> 0.95$, figure~\ref{FIG:MH125NN}.}, a factor of 10 improvement. 
A total of eleven artificial neural networks (NN)~\cite{Hocker:2007ht,Root} 
are used to improve the resolutions of variables sensitive to Higgs production 
and to separate the signal and background contributions. Altogether, the use 
of these NN leads to a 24\% increase in search
sensitivity\footnote{The search sensitivity is defined as the percentage reduction of the median expected limit.},
in addition to that expected from the inclusion of additional data with respect to the previous analysis~\cite{PhysRevD.84.052010}.

This analysis focuses on Higgs boson decays to \bbbar, and  
thus it is important to have the best possible resolution for \mbb. 
Section~\ref{SECTION:bJetEnergyCorrection} 
describes a NN used to correct the energies of $b$ jets, which in turn improves \mbb.
The untagged jets ($q$ jets) associated with each Higgs production
process have unique angular and kinematic distributions.
Section~\ref{SECTION:UntaggedJetsNeuralNetwork} describes three
networks that exploit these variables to identify $q$ jets from Higgs boson events. 
As gluon jets are typically wider than quark jets, jet width is useful 
for separating events containing
quark jets associated with Higgs-boson production from generic jets contained within
QCD multi-jet events, which are a mixture of quark and gluon jets. 
Section~\ref{SECTION:JetWidth} describes a technique for measuring
 jet width and a NN used to remove detector and kinematic dependences 
that also influence the jet width.

Section~\ref{SECTION:NNTraining} describes the final two-stage NN that 
is used to extract a potential signal contribution from the backgrounds.  The two-stage NN
can identify Higgs bosons produced by three different processes simultaneously. 
The first stage is based on three separate NNs trained
specifically to separate backgrounds from either {\it WH}, {\it ZH}, or VBF Higgs production, 
respectively, to exploit the unique characteristics of each signal process.
The outputs of the three process-specific NNs are used as inputs to a  second NN.   
The inputs to the first-stage networks are the corrected $b$-jet
energies, corrected $q$-jet widths, outputs of the $q$-jet networks,
and other  kinematic event variables.
In the previous search~\cite{PhysRevD.84.052010}, exclusive {\it VH} and VBF networks were used to search for Higgs bosons 
in non-overlapping signal regions.  The two-stage NN, developed for this search,
increases the search sensitivity by 15\%.
The use of a single signal region increases the number of potential Higgs boson
signal events by  20\%.
Both gains are above those expected from the inclusion of additional data alone.



%
\section{QCD multi-jet background prediction}~\label{SECTION:QCDMODEL}
Kinematic features of the QCD multi-jet background are predicted using a data-driven method.
An independent data control region is used to measure the probability for an event with one
$b$-tagged jet to contain an additional $b$-tagged jet (probe jet), referred to as the Tag Rate Function (TRF). 
%
%
%
The TRF is applied to data samples with exactly one jet $b$-tagged by SecVtx to
predict the distribution of events with two $b$-tagged jets.
The TRF is parameterized as a function of three variables:  
\et\, of the probe jet, 
$\eta$ of the probe jet, and $\Delta R$ between the tagged $b$ jet and probe jet, and 
implemented as a three-dimensional histogram.
The choice of variables used to parameterize the TRF is motivated by the kinematics 
of the QCD multi-jet background and the characteristics of the $b$-tagging algorithms. 
For example, the production of $b$ jets from gluon splitting has a  different $\Delta R$ distribution compared to 
direct production, and the probe jet \et\, and $\eta$
expresses aspects of the $b$-tagging algorithms and QCD multi-jet production. 
Further information on the technique can
be found in~\cite{PhysRevLett.103.221801}.


%
%

\begin{figure}[htbp]
  \centering
  \includegraphics[width=10cm]{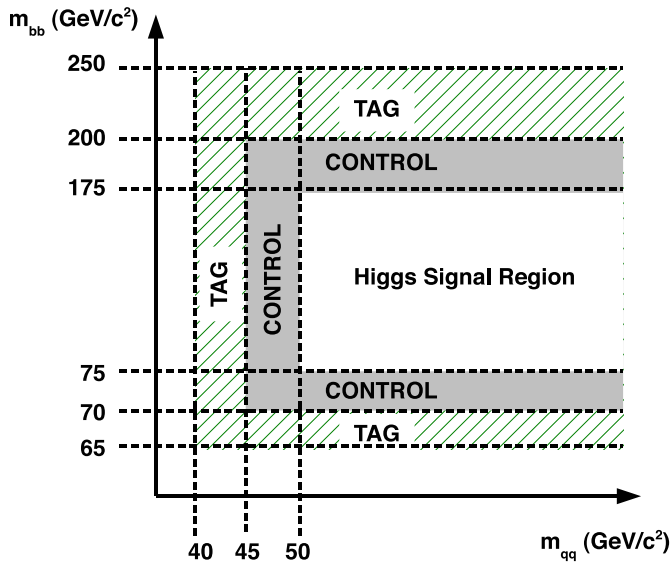}
  \caption{Signal and controls regions in the \mbb-\mqq\, plane. The \Tag\, region
    is used to derive the TRF for modeling the QCD multi-jet background. 
    The  \Ctrl\, region is used to test and     
     derive systematic uncertainties of this background model.}
  \label{FIG:MbbMqqRegion}
\end{figure}

We use separate TRFs for SS and SJ events, which are obtained from
events in the \Tag\, region (figure~\ref{FIG:MbbMqqRegion}), 
defined as $\mqq \in [40\,\gevcc, 45\,\gevcc]\, \cup\, \mbb \in [65\,\gevcc,
250\,\gevcc]$ and $\mqq > 45\, \gevcc \cup \mbb \in [65\,\gevcc, 70\,\gevcc]\, \cup
\mbb \in [200\,\gevcc, 250\,\gevcc]$.
To validate the background model, the TRF is tested in the 
TAG (for self-consistency) 
and two other control regions non-overlapping with
the signal region (figure \ref{FIG:MbbMqqRegion}):
the \Ctrl\, region, defined as $\mqq \in [45\,\gevcc, 50\,\gevcc] \cup
\mbb~\in~[70\, \gevcc,
200\,\gevcc]$ and $\mqq>50\, \gevcc \cup \mbb \in [70\,\gevcc, 75\,\gevcc] \cup
\mbb \in [175\,\gevcc, 200\,\gevcc]$;
and the \Njet\, control region defined as sharing the
same \mbb\, and \mqq\, criteria as the signal
region, but contains those events with six reconstructed jets.  
The TRF prediction of different variables are compared to data in these control regions
and any shape difference is propagated as an uncertainty of the QCD multi-jet model.


The \mqq\, variable is not perfectly modeled by the TRF. The residual 
mismodeling is corrected by following the
procedure defined in previous searches~\cite{PhysRevLett.103.221801,PhysRevD.84.052010},
which reweights events as a function of the observed \mqq.
The correction function is derived from a
fit to the ratio of the observed \mqq\, over the same quantity
predicted by TRF in events from the \Tag\, region.

Figures~\ref{FIG:TrainingVariables_SS_1}-\ref{FIG:TrainingVariables_SS_4}
show a comparison of observed data and background predictions
in the signal region for the variables used in the final signal discrimination neural network
(section~\ref{SECTION:NNTraining}) after application of the  \mqq\, correction function.
The modeling of some variables appears to be poor but the differences are within the 
shape uncertainties of the QCD multi-jet prediction. 
More details on these variables are given in the following sections.

\section{Energy correction for $b$ jets}~\label{SECTION:bJetEnergyCorrection}
The experimental resolution of the invariant mass of the two $b$ jets, \mbb, 
has a significant effect on the sensitivity of our search. 
To improve the \mbb\, resolution, a NN is trained to estimate the correction factor required to obtain the best 
possible estimate of the parent $b$-parton energy from the measured jet energy~\cite{bjet}.



A NN is trained for each $b$-tagging algorithm.
Nine variables, describing a given jet, are used to train the NN for
SecVtx tagged jets. These are the jet \et, the jet
transverse momentum ($\pt \equiv p \sin \theta$),
the \et\, before the application of jet energy
correction 
(uncorrected jet \et), the transverse mass\footnote{The
  transverse mass
  is defined as $(\pt/p)M$, where $M$ is the invariant mass of the jet.}, the decay
length of the jet
in the transverse plane\footnote{The decay length is defined as the
  transverse distance between the primary vertex and the
reconstructed secondary vertex in the SecVtx $b$-tagged jet.} and
its uncertainty, the \pt\, of the secondary vertex, the maximum
\pt\, of the tracks inside the jet cone, and the \pt\, sum of all tracks within
the jet cone. Six variables are used to train the NN for JetProb
tagged jets: the jet \et, the jet
\pt, the uncorrected jet \et, the transverse mass, 
the maximum \pt\, of the tracks inside the jet cone, and the \pt\, sum
of all tracks within the jet cone.

The NNs are trained using simulated VBF events\footnote{All NNs in this
  paper are trained using statistically independent samples.} with Higgs masses from
100\,\gevcc\, to 150\,\gevcc\, at 5\,\gevcc\, intervals. Events are required
to pass the selection described in section \ref{SECTION:EventSelection} and each $b$-tagged jet is
required to be matched geometrically with a $b$ parton. 
The matching criterion requires the
$\Delta R$ between the $b$ jet and $b$ parton not to exceed 0.4.
SecVtx- and JetProb-tagged jets are used to train the SecVtx and JetProb networks, respectively.  

\begin{figure}[htbp]
  \centering
  \includegraphics[width=10cm]{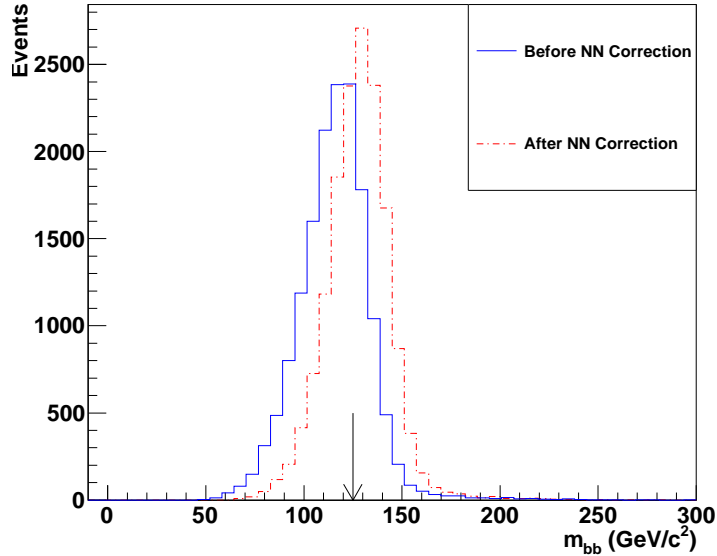}
  \caption{Comparison of \mbb\, distribution in simulated decays of
    $125\,\gevcc$ Higgs bosons produced through VBF, before and after the $b$-jet
    energy correction for a VBF MC sample with $m_H=125\,\gevcc$
    (indicated by the black arrow).}
  \label{fig:mass125}
\end{figure}

Figure \ref{fig:mass125} shows the \mbb\, distribution in simulated
decays of $125\,\gevcc$ Higgs bosons produced through VBF, before
and after $b$-jet energies are corrected.  The mean shifts from $116\,\gevcc$
to $128\,\gevcc$ and the root mean square (RMS) from $15.6\,\gevcc$ to
$13.7\,\gevcc$.
The resolution, defined as the ratio between the RMS and the mean,
shifts from $0.13$ to $0.11$, an improvement of $18\%$.

The $b$-jet energy corrections should be independent of the sample
used to train and test the NN. The NN training and testing was
repeated using {\it WH} and {\it ZH} events and similar results were obtained.

\section{Untagged jets neural network}~\label{SECTION:UntaggedJetsNeuralNetwork}
The angular distributions of untagged jets ($q$ jets) from {\it VH} or VBF
differs from the angular distributions of generic jets contained within QCD multi-jet background events.
Identification of $q$ jets can therefore help to separate signal events from QCD
multi-jet background contributions. 
In particular, the \mqq\, obtained from $q$ jets associated with the {\it WH} and {\it ZH} processes 
is constrained by the mass of the $W$ and $Z$, respectively. The $q$ jets produced in VBF events
are typically separated by large $\phi$ and $\eta$, while the $q$ jets in QCD multi-jet events 
tend to exhibit a large difference in $\phi$ and a small difference in $\eta$. Three networks~\cite{Hocker:2007ht}, 
referred to as 
{\it qq\_\!WH} NN, {\it qq\_ZH} NN, and {\it qq\_}\!VBF NN, are trained to separate events with $q$ jets
originating 
from {\it WH}, {\it ZH}, and VBF production from background events. 
The input variables are
$m_{qq}$, $\Delta \phi_{qq}$, $\Delta \eta_{qq}$, $\Delta R_{qq}$, and the transverse momenta of each
$q$ jet with respect to the total momentum of the system. The networks are trained using Higgs MC to model
signal and data-driven prediction for QCD multi-jets to model
background.
Examples of the output distributions of these trained neural network are shown in figure~\ref{FIG:TrainingVariables_SS_5}.

\section{Jet width}~\label{SECTION:JetWidth}
The untagged jets ($q$ jets) associated with the QCD multi-jet background
are a mixture of quark and
gluon jets whereas the $q$ jets associated with the Higgs signal are predominantly quark jets. 
As gluon jets tend to be broader than quark jets, jet width
is another useful variable for distinguishing potential
Higgs events from the background.
We defined jet widths measured within the calorimeter 
($\langle R \rangle_{\rm CAL}$) and tracker ($\langle R \rangle_{\rm TRK}$) as 

\begin{eqnarray}
\langle R \rangle_{\rm CAL} = \sqrt{ {\Large \sum_{\text{towers}}}  \left[
    \frac{ E_t^{\text{tower}} }{ E_t^{\text{jet}}  } \;\Bigl( \Delta R
    (\textrm{tower,jet}) \Bigr)^{2} \right]  } \\
\langle R \rangle_{\rm TRK} = \sqrt{ {\Large \sum_{\text{tracks}}}  \left[
    \frac{ P_t^{\text{track}} }{ P_t^{\text{jet}}  } \;\Bigl( \Delta R
    (\textrm{track,jet})   \Bigr)^{2} \right]  },
\label{EQN:JetShapeDefinition}
\end{eqnarray}

\noindent
where $\Delta R$(tower,jet) ($\Delta R$(track,jet)) is the angular
distance between the jet axis and the calorimeter tower (track).
All calorimeter towers within the jet cone of $\Delta R < 0.4$ are
used in the $\langle R \rangle_{\rm CAL}$ calculation. 
All tracks with $\pt>1$\,\gevc\, and within the jet cone of $\Delta R < 0.4$ are 
used in the calculation of $\langle R \rangle_{\rm TRK}$.

The jet width also  
varies as a function of 
jet~\et, jet $\eta$, and the number of primary vertices ($N_{\rm vtx}$),
and is parameterized by a neural network fit.
%
%
%
%
These dependences are removed by rescaling the measured
jet widths to a common reference (that for a jet with
\et=50\,GeV, $\eta$=0, and $N_{\rm vtx}$=1) using the procedure described in ref.~\cite{PhysRevD.84.052010}.  
%
%
%
%
%
%
%
%
%
The NN function to parameterize the variation of jet width with jet~\et, jet $\eta$, and $N_{\rm vtx}$,
is trained on a sample of untagged quark jets
from the hadronic $W$ boson decays in $t\bar{t} \rightarrow \bbbar
l\nu qq$ ($\ell=e,\mu$) events. 
%
%
The highest \et\, untagged-jet pair whose  invariant mass is in the
range $50 - 110\,\gevcc$
is assumed to originate from the hadronic $W$ boson decay.
%
%
Separate networks are trained for MC and data.
After rescaling, any differences in the jet width are assumed to be associated with
the type of parton that initiated the jet.
%
The \ttbar\, MC and data $q$-jet width distributions are found to agree after rescaling the measured jet widths. 
To check that the jet width rescaling can be applied to non-\ttbar\,
samples, the rescaling is also applied to the $q$ jets in {\it WH}, {\it ZH}, and VBF MC events. 
The mean rescaled jet width in all samples is consistent  with the
width observed in the 
\ttbar\, sample, which verifies the independence of the corrections with respect to jet \et, $\eta$, and $N_{\rm vtx}$.


A systematic uncertainty is assigned by adding an offset to the
rescaled \ttbar\, MC jet width and comparing the $\chi^{2}/\text{degree of freedom}$ ($\chi^{2}/{\rm d.o.f}$) of the shifted
MC and \ttbar\, data distributions with the unshifted MC and data.  The
uncertainty is defined by the offset that changes the $\chi^{2}/{\rm d.o.f}$
by $\pm 1$ unit.
The calorimeter jet width uncertainty is
$\pm 2.6\%$ and the tracker jet width uncertainty is $\pm 5.5\%$.

Figures~\ref{FIG:TrainingVariables_SS_2}\subref{width_tower_q1}-\ref{FIG:TrainingVariables_SS_2}\subref{width_tracks_q2} show the corrected 
jet width distributions of the untagged jets measured by the calorimeter and tracker. The Higgs signal tends 
to lower jet width values, which implies quark-like, whereas the QCD multi-jet tends to higher jet width, which
implies a mixture of quark and gluons. The jet width distributions of the Higgs signal is different to the background
which shows this variable is useful for the Higgs boson search.


\section{Classification of Higgs boson events}~\label{SECTION:NNTraining}
A final NN is trained to optimize the separation of signal and background~\cite{Hocker:2007ht}, which
incorporates information from kinematic and angular jet variables, jet widths, event shapes, and the outputs of the untagged jets
 ($q$ jets) NNs. 
 %
 %
%
%
The energies of the $b$ jets and widths of the $q$ jets 
are corrected as described in sections~\ref{SECTION:bJetEnergyCorrection} and~\ref{SECTION:JetWidth},
respectively. 
%
%
%
%
As the {\it WH}, {\it ZH}, and VBF processes have different
kinematics, 
dedicated {\it WH}, {\it ZH}, and VBF networks are trained separately
for each process, resulting in three outputs. 
The outputs of the process-specific NNs are combined as inputs to a grand NN, 
referred to as the \SD.  
The output of the \SD\, is used to obtain Higgs search limits. 

The selection of input variables for the
process specific {\it WH}, {\it ZH}, and VBF networks training must fulfill two
criteria: the variables must have good background-to-signal separation,
 and they must be well modeled by TRF.
%
%
The discriminating variables for
the {\it WH}-NN and {\it ZH}-NN training are $\mbb$, $\mqq$, the cosine of the leading-jet
scattering angle in the four-jet rest-frame ($\cos( \theta_{3}) $)~\cite{PhysRevD.53.4793}, the $\chi$
variable\footnote{$\chi$ variable is the minimum of $\chi_{W}$ and
  $\chi_{Z}$ where $\chi_{W}   =  \sqrt{(M_{W} - \mqq)^2 + (M_H -
    \mbb)^2}$ and a similar expression exists for
  $\chi_{Z}$.}~\cite{PhysRevD.84.052010}, the calorimeter jet width of
the first ($\langle R \rangle^{q_1}_{CAL}$) and second leading untagged
jet ($\langle R \rangle^{q_2}_{CAL}$), the tracker jet width
of the first  ($\langle R \rangle^{q_1}_{TRK}$) and second leading
untagged jet ($\langle R \rangle^{q_2}_{TRK}$), 
%
%
aplanarity\footnote{Aplanarity measures the transverse momentum
  component out of the event plane.}, sphericity\footnote{Sphericity is
a measure of the summed transverse momentum squared
with respect to the event axis.}, centrality\footnote{Centrality
measures how much of the energy flows into the central rapidity region.}~\cite{Sjostrand:2000wi}, 
$\Delta$R of the two $b$-tagged jets,  
$\Delta$R of the two untagged jets, $\Delta\phi$ of the two
$b$-tagged jets, 
$\Delta\phi$  of the two untagged jets, 
and the {\it qq\_\!WH} and {\it qq\_ZH} network outputs
(section \ref{SECTION:UntaggedJetsNeuralNetwork}).
Not all variables used in the {\it WH} and {\it ZH} networks' training have a good
background-to-signal separation for VBF.
For the VBF-NN
training, the $\cos( \theta_{3}) $, the aplanarity, and
the $\Delta\phi$ of the two untagged jets are removed; the $\eta$ angle
of the first ($\eta_{q_1}$) and second leading untagged jet ($\eta_{q_2}$), 
the $\Delta\eta$ of the two untagged jets ($\Delta\eta_{qq}$),
the invariant mass of four jets
system, the sum of the four jets' momenta along $z$ direction are added,
and the {\it qq\_\!WH} and {\it qq\_ZH} network outputs are replaced by {\it qq\_}\!VBF NN output. 
%
%
Overall, the {\it VH}(VBF)-NN is trained with 17(18) variables, of which \mbb\,~and~\mqq\, (\mqq\,~and~$\Delta\eta_{qq}$)  are the most discriminating variables.

\begin{figure}[htbp]
  \begin{center}
    \subfigure[]{\label{mass_bb}\includegraphics[width=5.5cm]{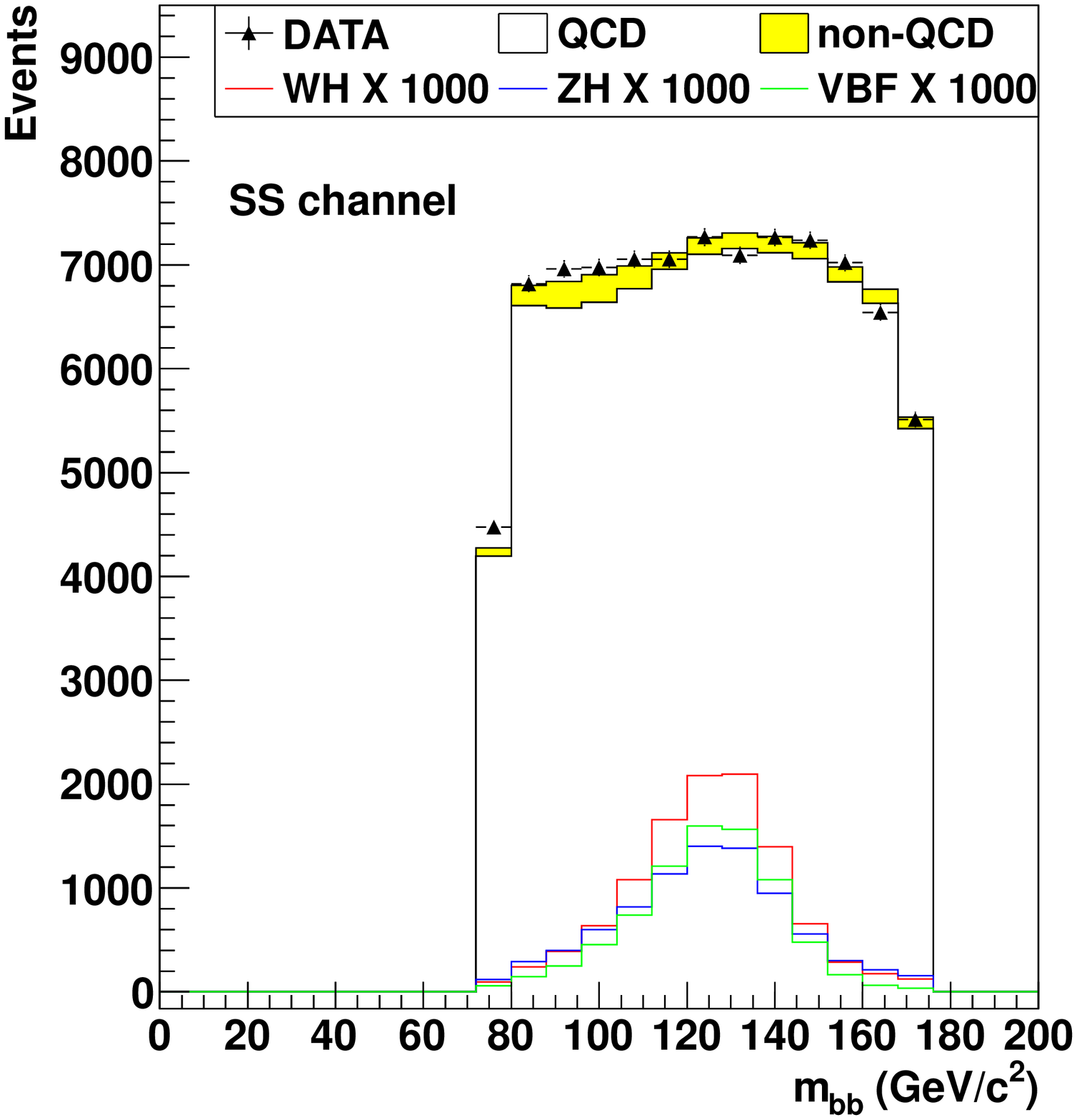}}
    \subfigure[]{\label{mass_qq}\includegraphics[width=5.5cm]{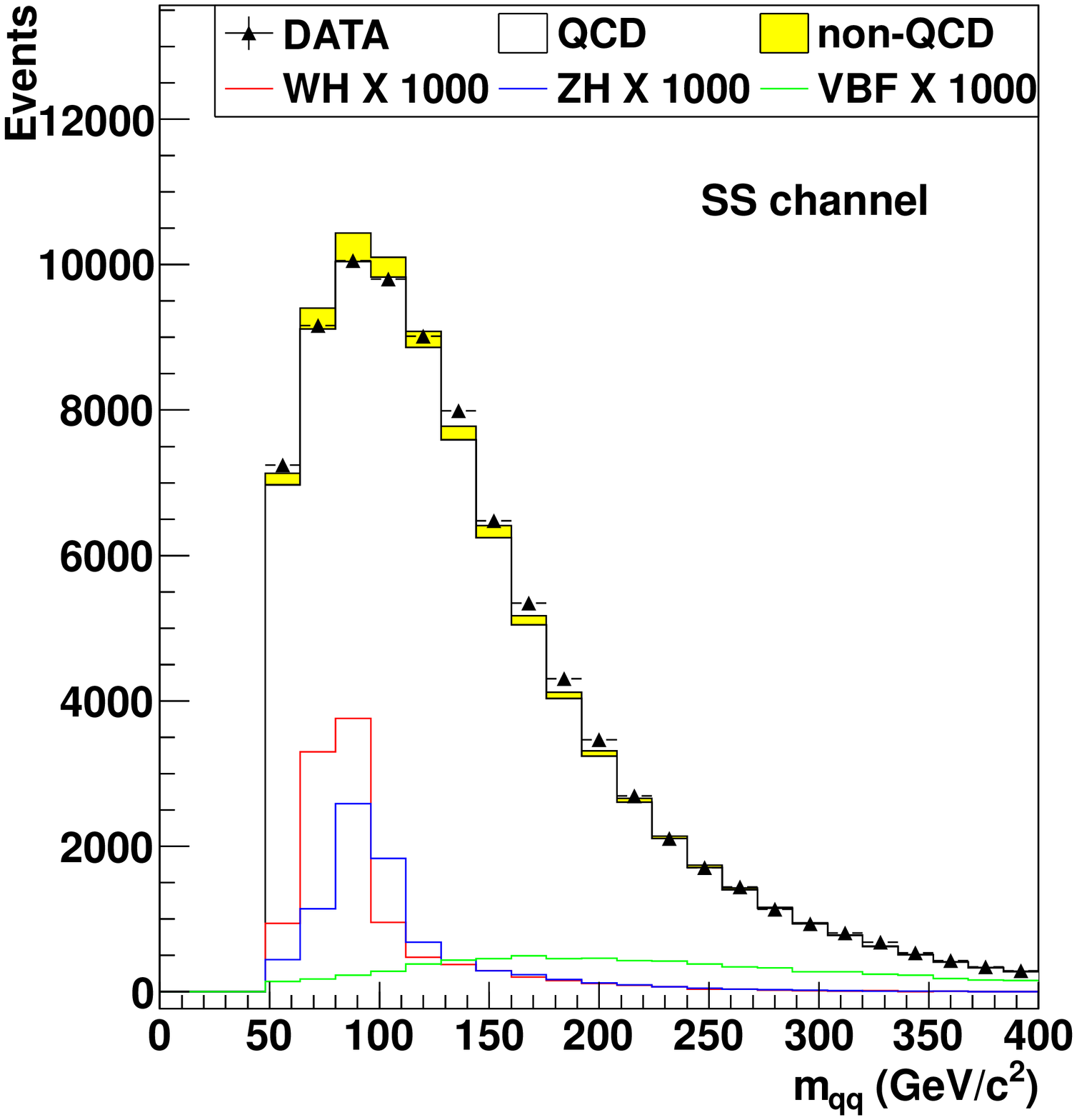}} \\
    \subfigure[]{\label{mass_4j}\includegraphics[width=5.5cm]{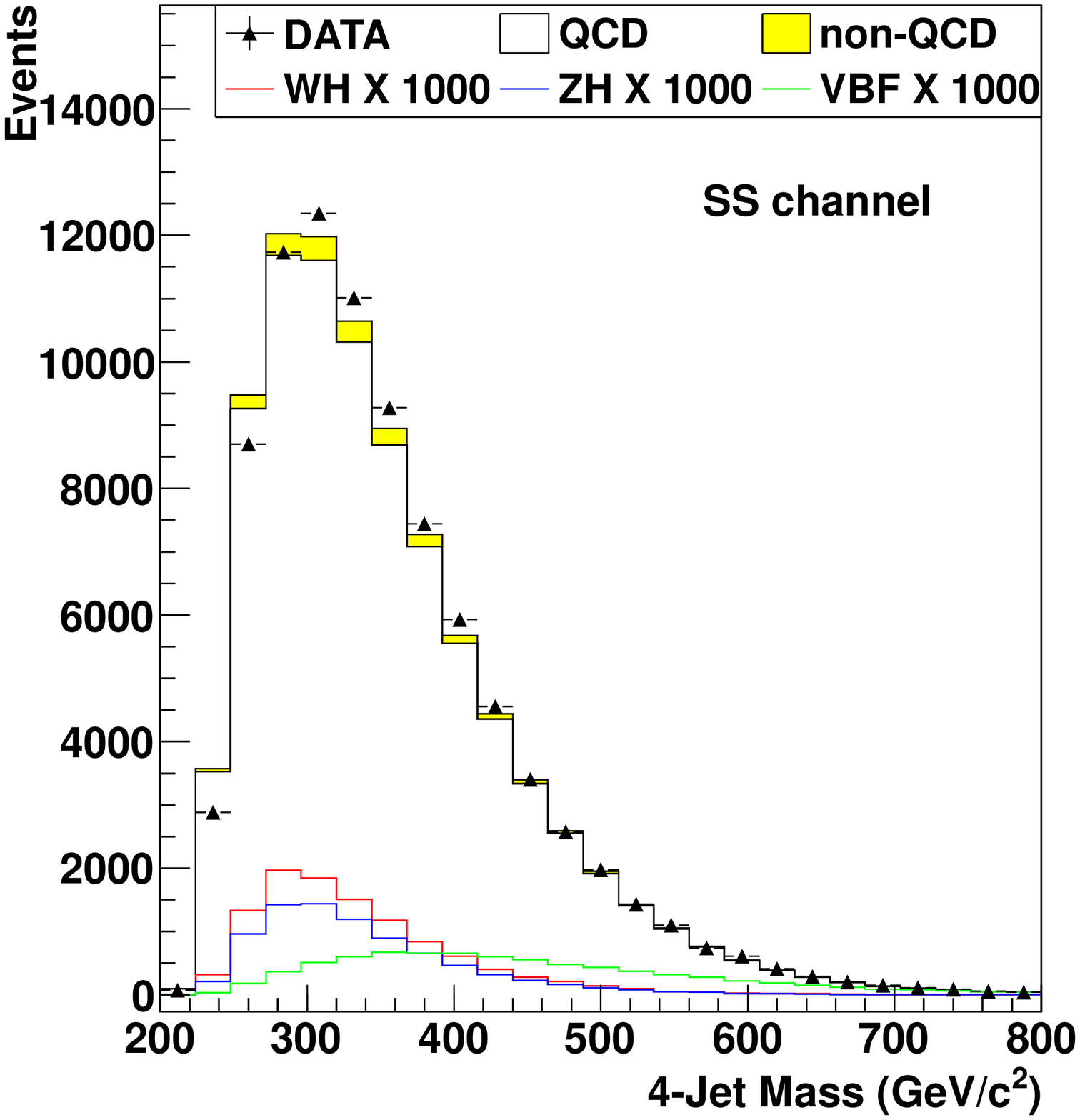}}
    \subfigure[]{\label{sumPz_4j}\includegraphics[width=5.5cm]{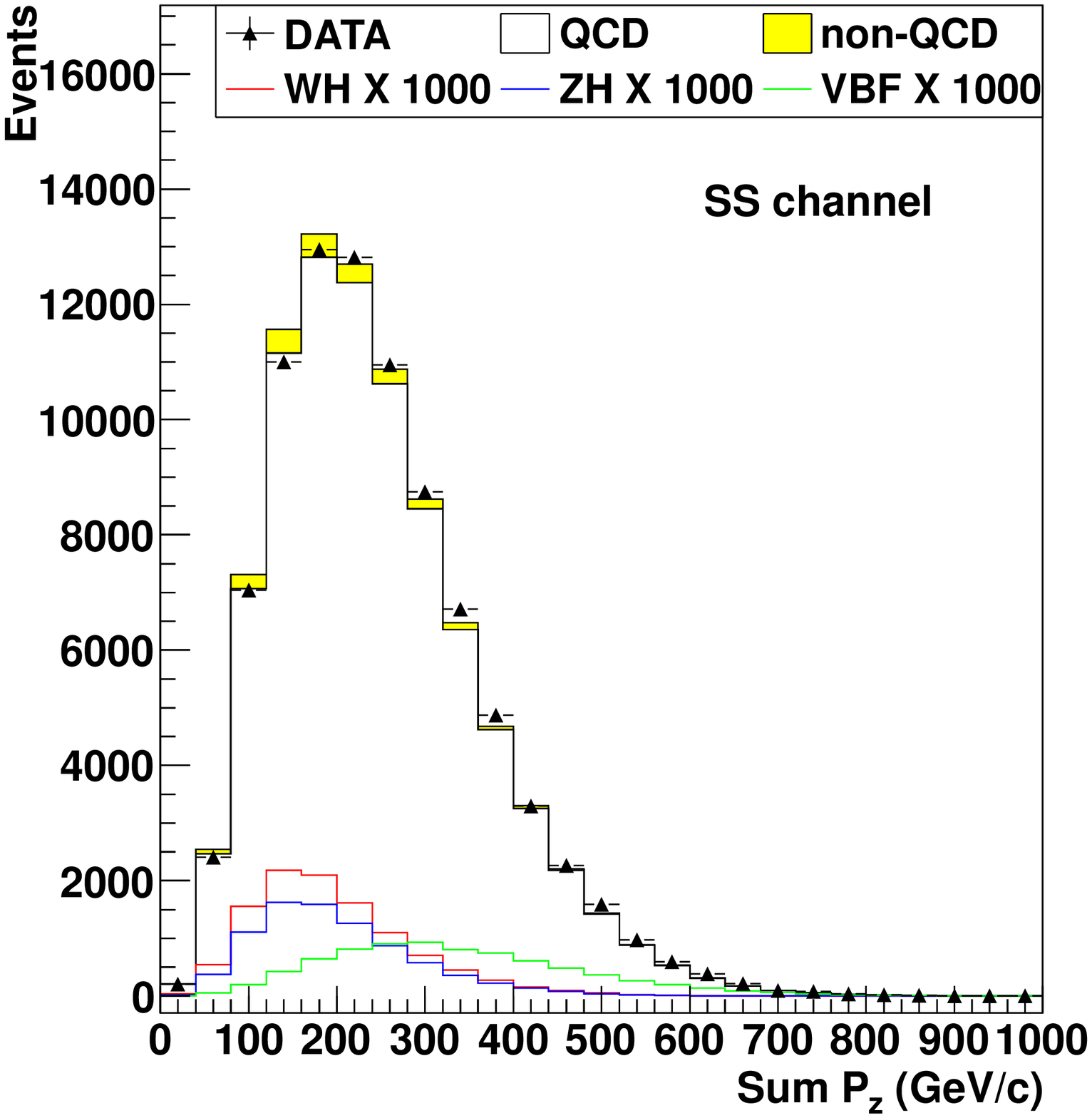}} \\
    \caption{The QCD multi-jet background prediction (SS $b$-tag category) for
      \subref{mass_bb} \mbb,
     \subref{mass_qq} \mqq, \subref{mass_4j} the invariant mass of four-jets
      system, and \subref{sumPz_4j} the sum of the
      momenta along $z$ direction for each of the four jets in the search signal region. 
      The \mqq\, variable distribution is obtained after
      the application of the \mqq\, correction described in section~\ref{SECTION:QCDMODEL}.
      The black histograms are the TRF derived predictions for the QCD multi-jet background, and the black triangles
      are the data. 
      The yellow histogram shows the MC predicted 
      non-QCD background which is the sum of 
      \ttbar, single-top, $Z+$jets, $W+HF$, and diboson contributions.
	The predicted distributions for {\it WH} events (red), {\it ZH} events (blue), and VBF events (green) for
	a Higgs mass of $m_{H}=125\,\gevcc$ scaled by a factor of 1000 are also shown. 	
      } 
    \label{FIG:TrainingVariables_SS_1}
  \end{center}
\end{figure}

\begin{figure}[htbp]
  \begin{center}
    \subfigure[]{\label{cos_theta_3}\includegraphics[width=5.5cm]{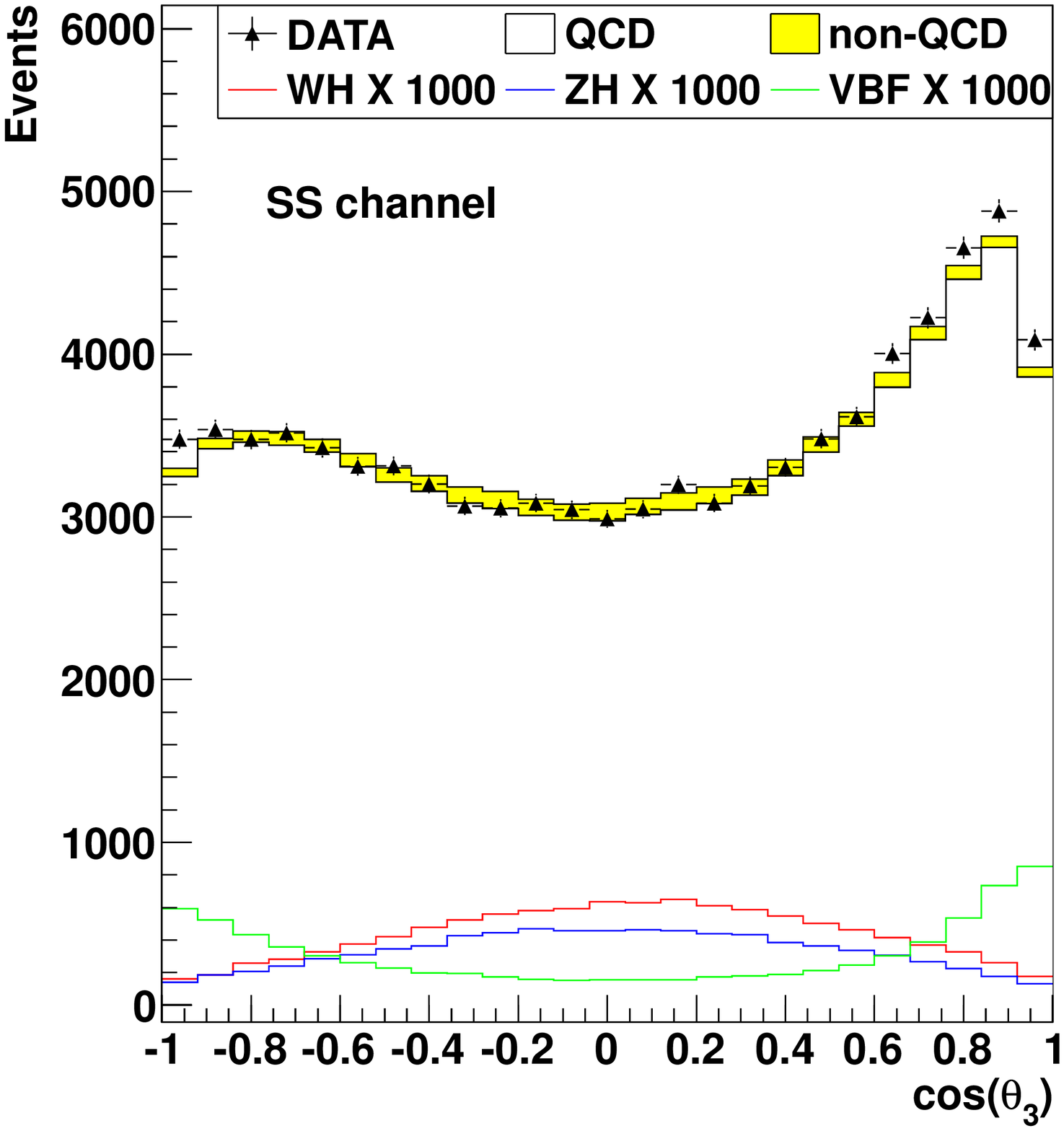}} 
    \subfigure[]{\label{chi125}\includegraphics[width=5.5cm]{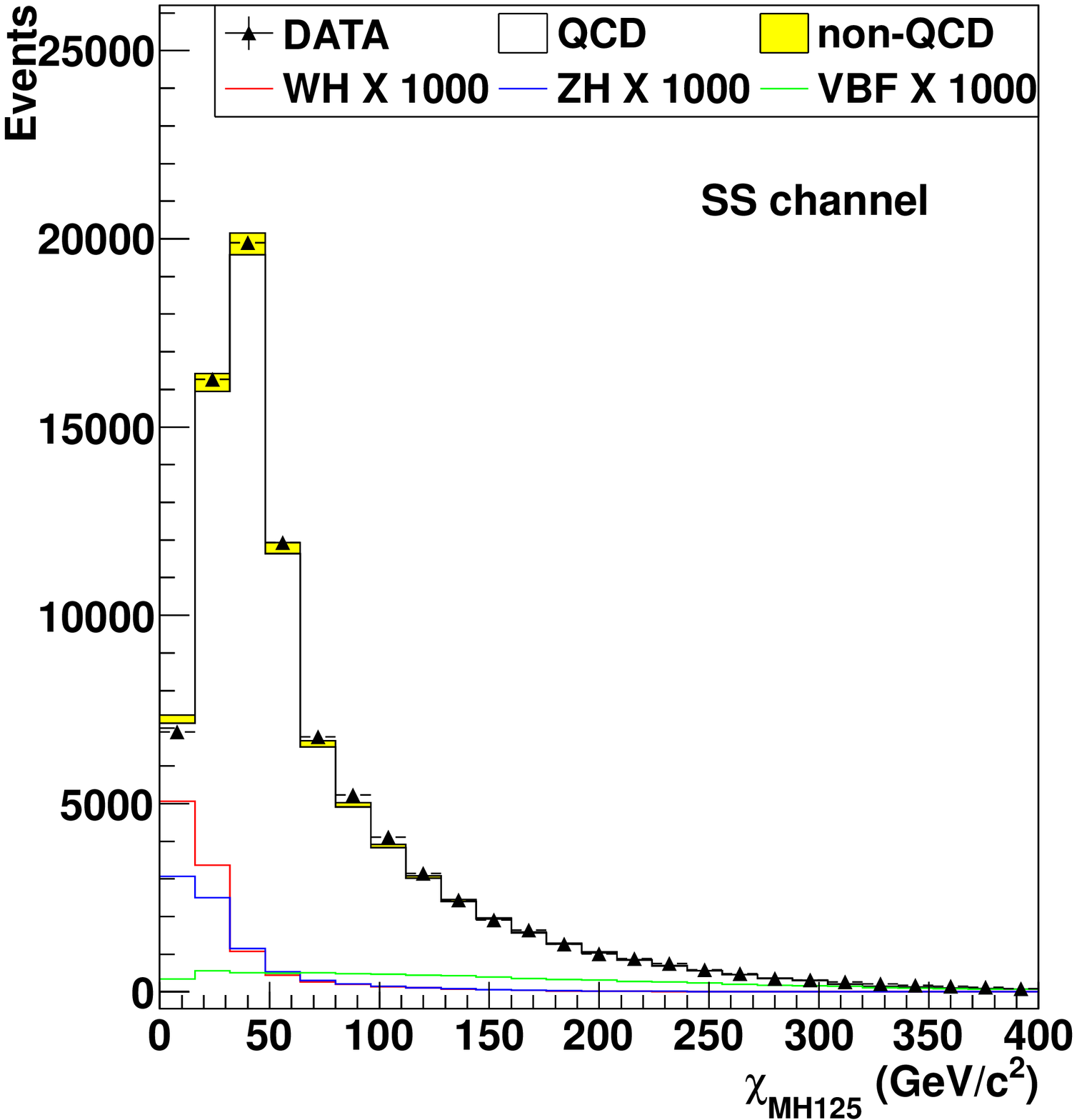}} \\
    \subfigure[]{\label{width_tower_q1}\includegraphics[width=5.5cm]{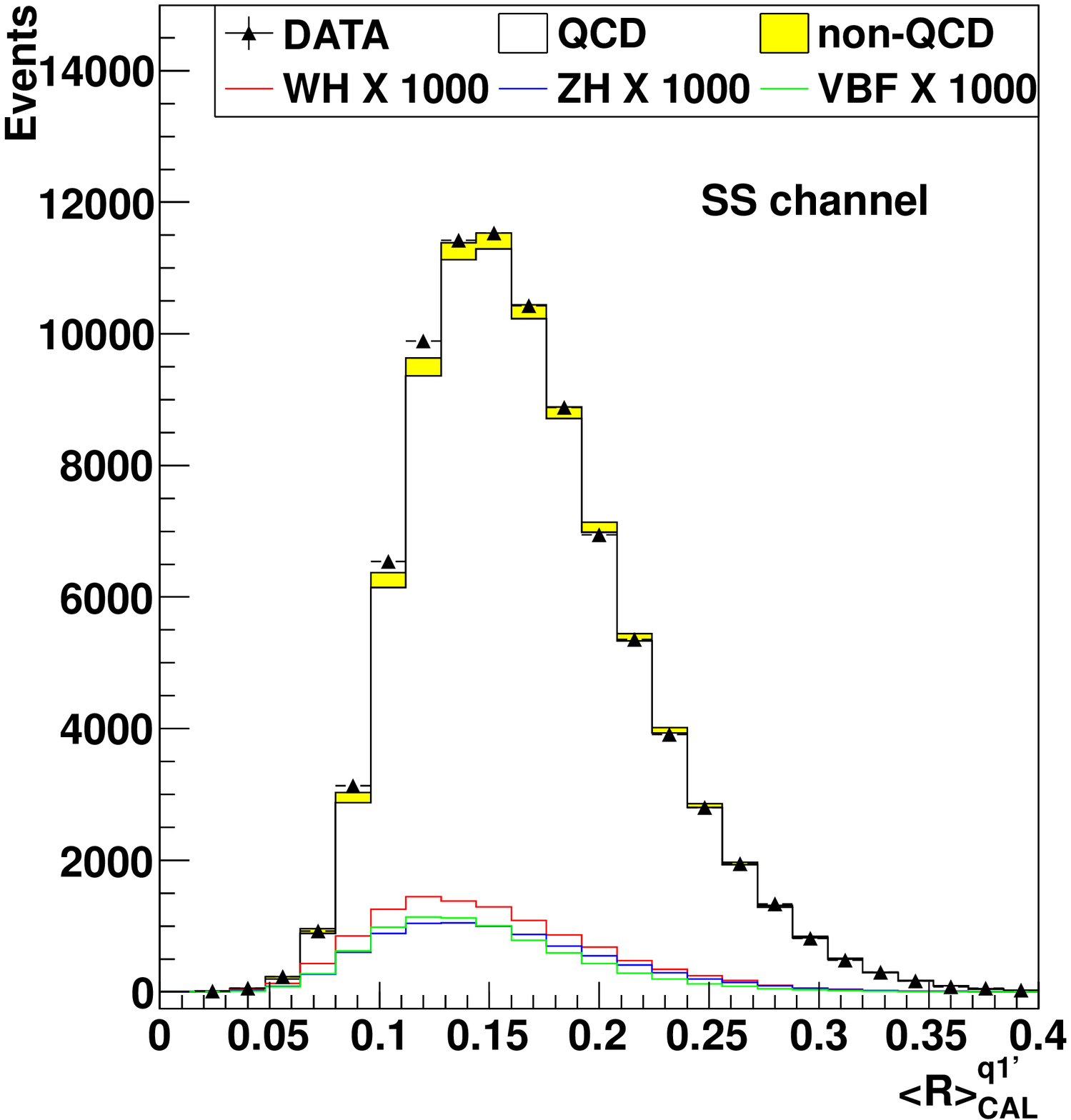}}
    \subfigure[]{\label{width_tower_q2}\includegraphics[width=5.5cm]{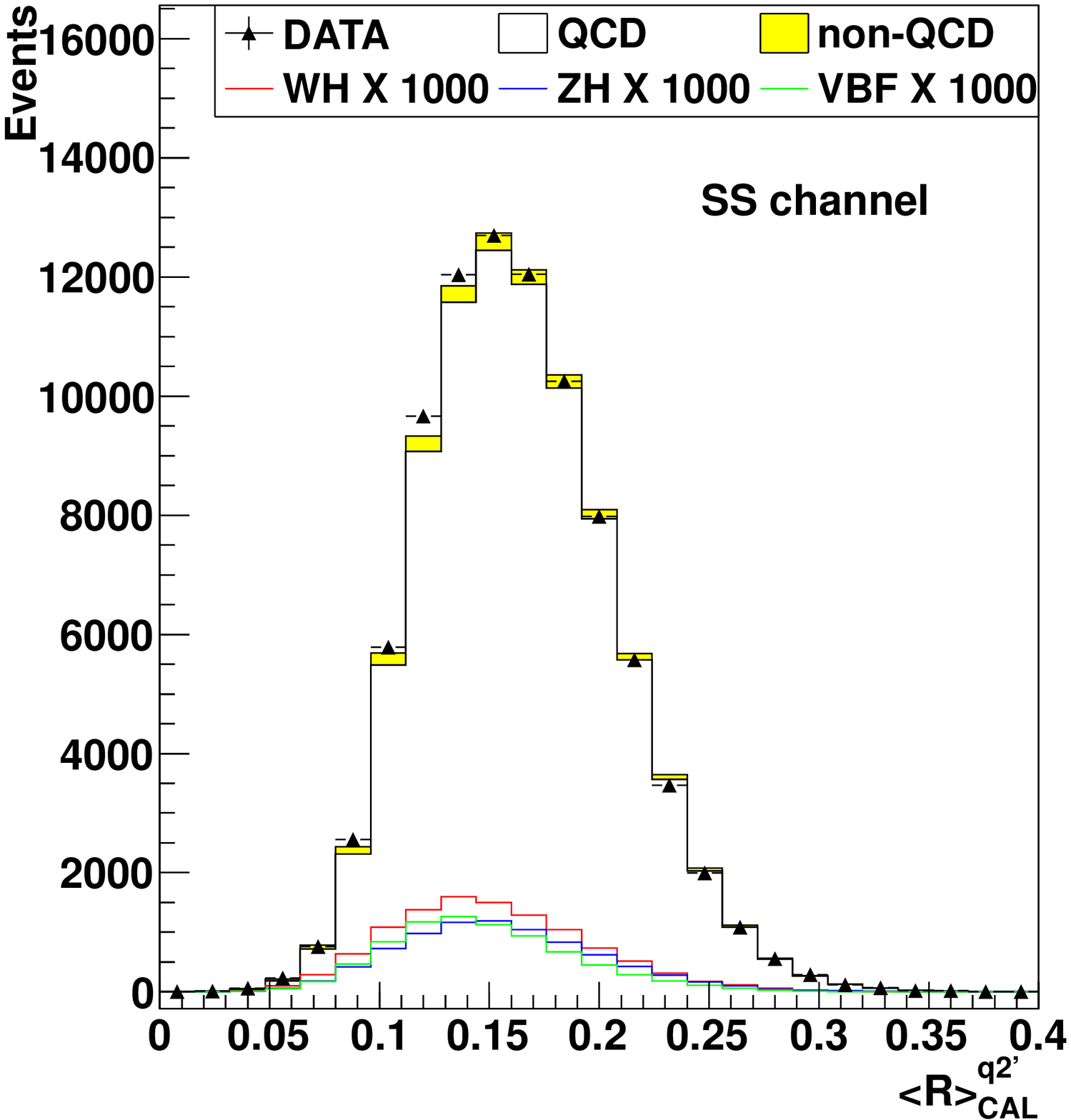}}
        \subfigure[]{\label{width_tracks_q1}\includegraphics[width=5.5cm]{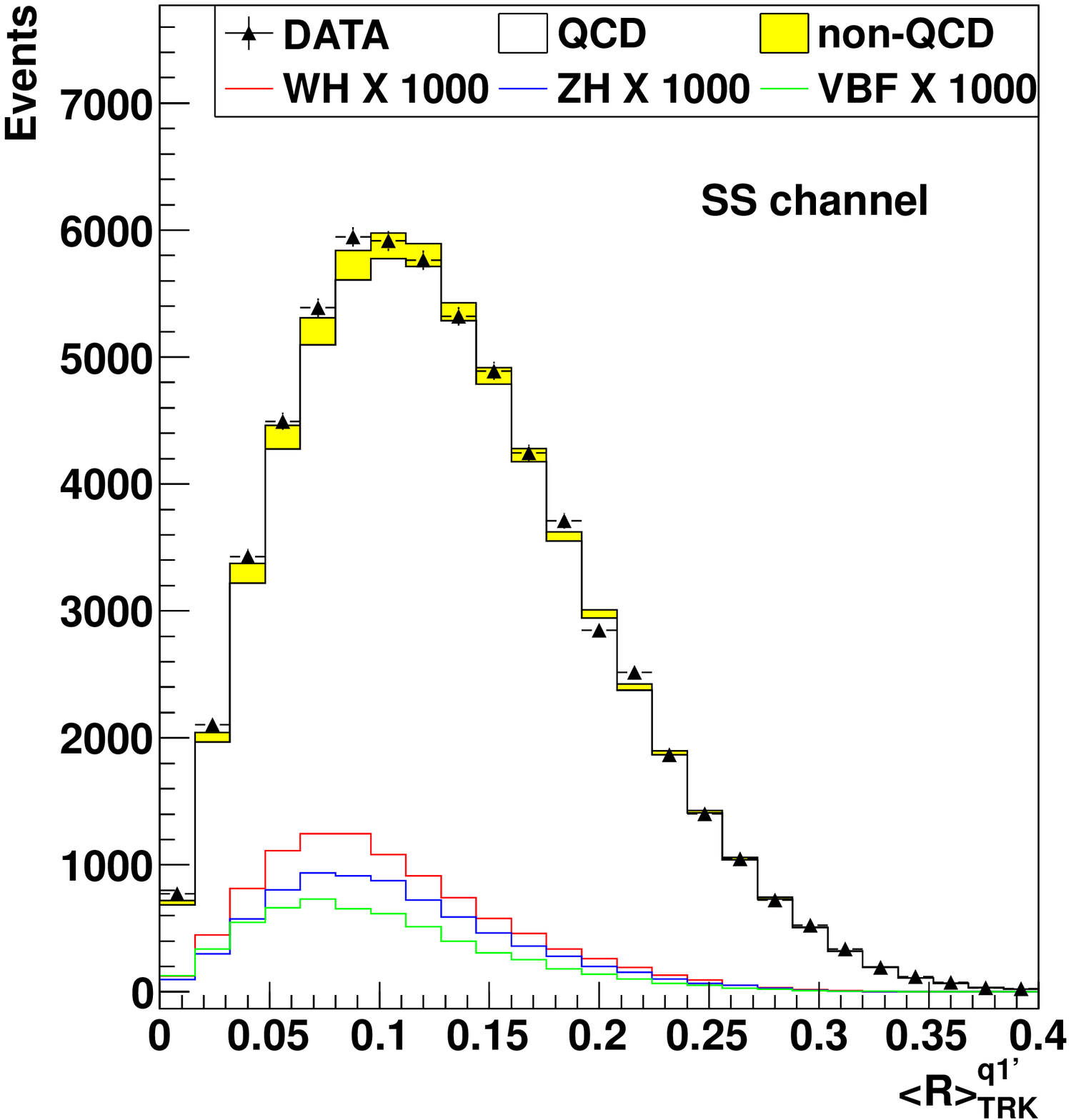}}
    \subfigure[]{\label{width_tracks_q2}\includegraphics[width=5.5cm]{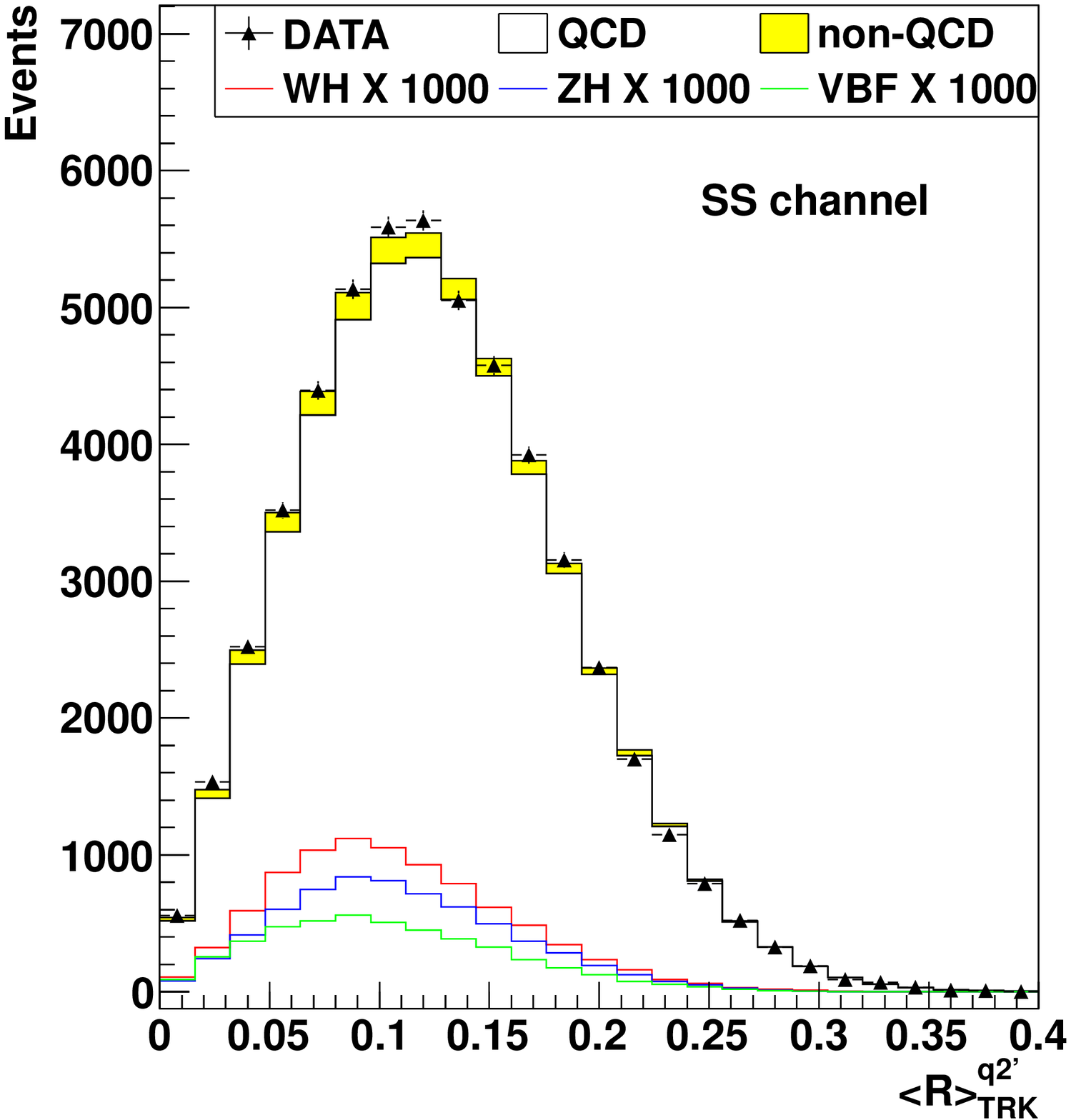}} \\
    \caption{The QCD multi-jet background predictions for the SS $b$-tag category of 
    \subref{cos_theta_3} the cosine of the leading-jet scattering angle in the four-jet rest-frame~\cite{PhysRevD.53.4793},
      \subref{chi125} the $\chi$ variable~\cite{PhysRevD.84.052010},  
      \subref{width_tower_q1} the calorimeter jet width of the first and \subref{width_tower_q2} second leading untagged jet, and    
     \subref{width_tracks_q1} the tracker jet width of the first and \subref{width_tracks_q2} second leading untagged jet.
     Descriptions of the signal and background histograms can be found
     in the caption of figure~\ref{FIG:TrainingVariables_SS_1}.}
     
    \label{FIG:TrainingVariables_SS_2}
  \end{center}
\end{figure}

\begin{figure}[htbp]
  \begin{center}
     \subfigure[]{\label{eta_q1}\includegraphics[width=5.5cm]{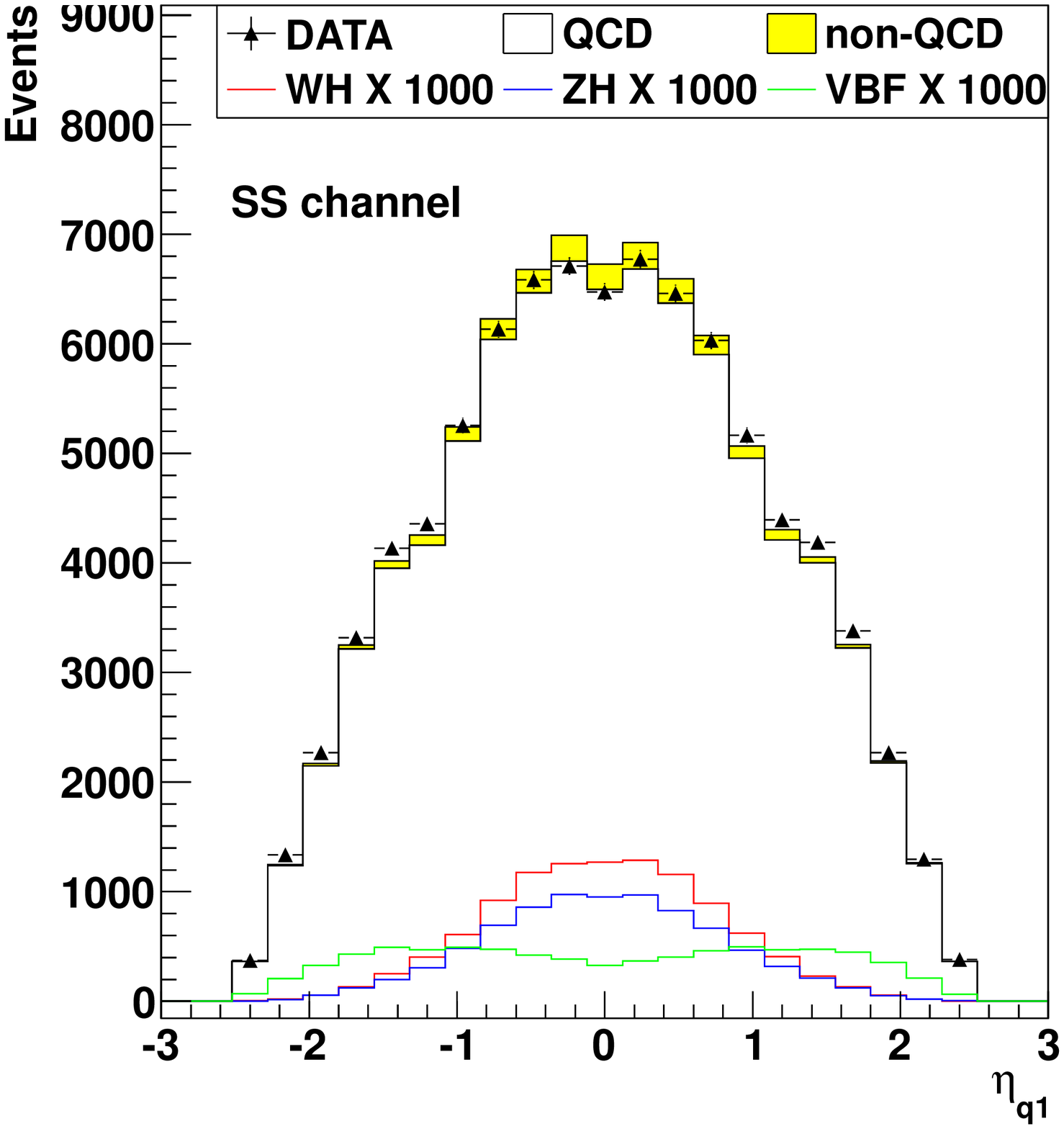}} 
     \subfigure[]{\label{eta_q2}\includegraphics[width=5.5cm]{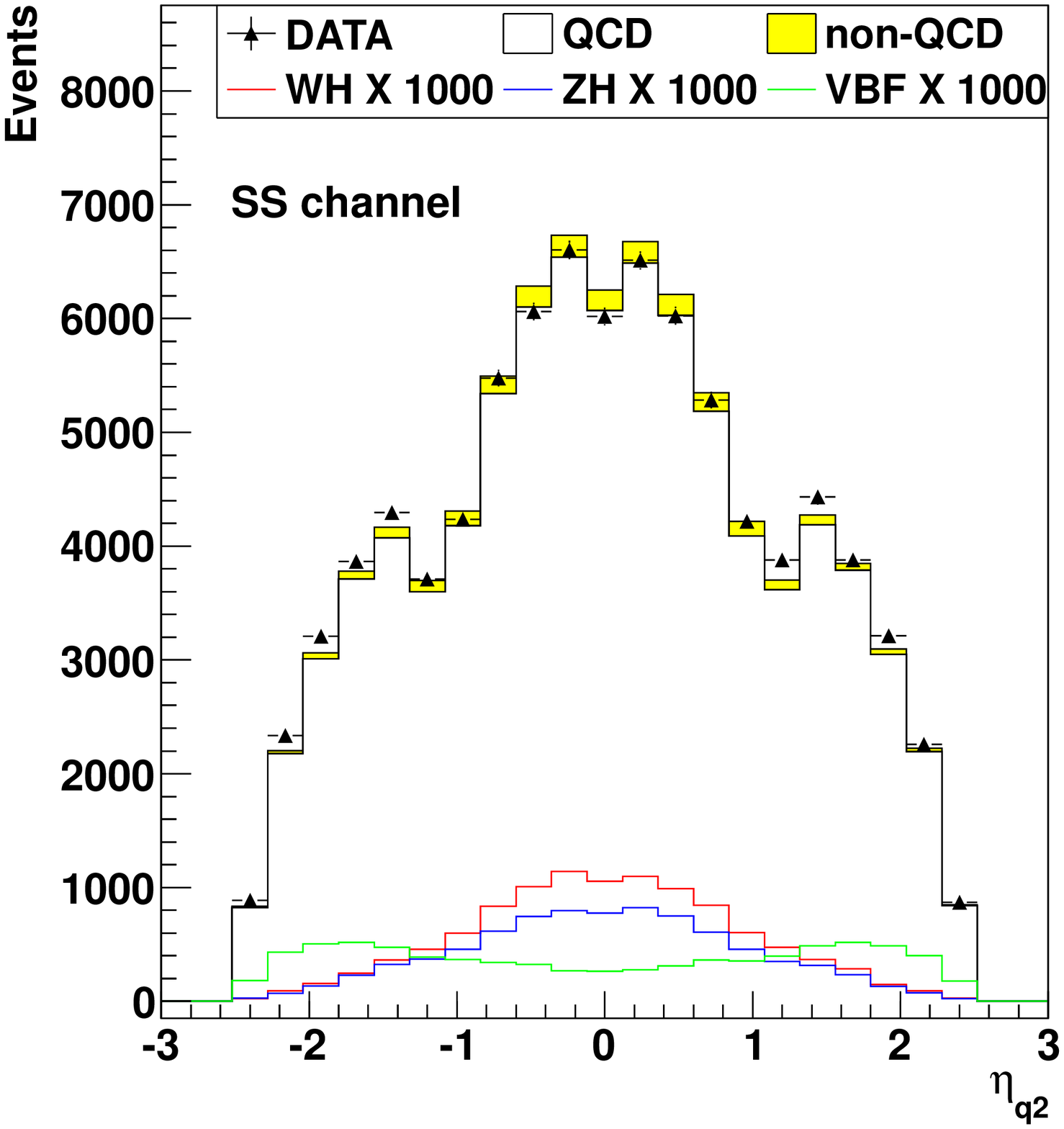}} \\
    \subfigure[]{\label{delta_eta_qq}\includegraphics[width=5.5cm]{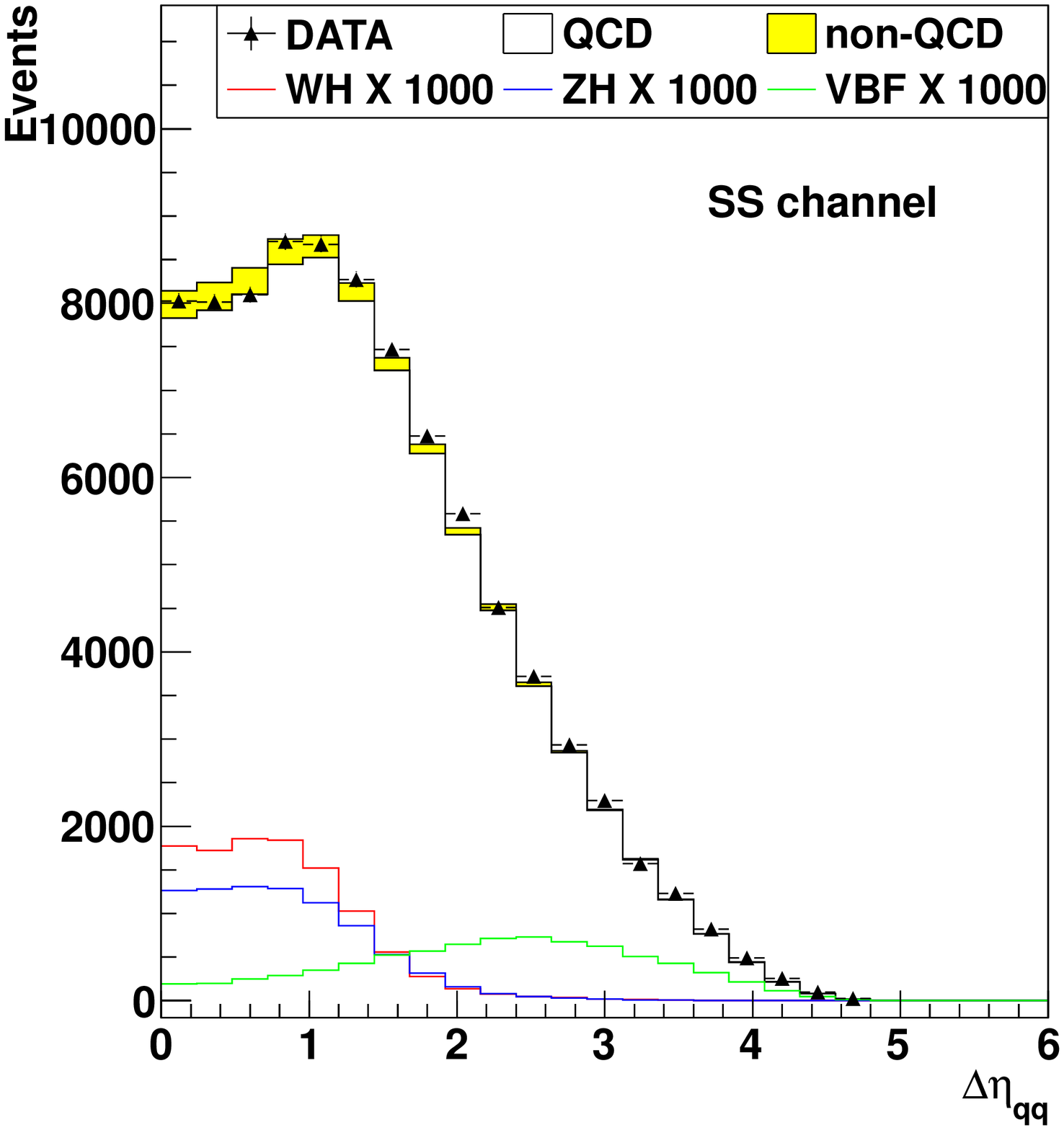}}
    \subfigure[]{\label{aplan}\includegraphics[width=5.5cm]{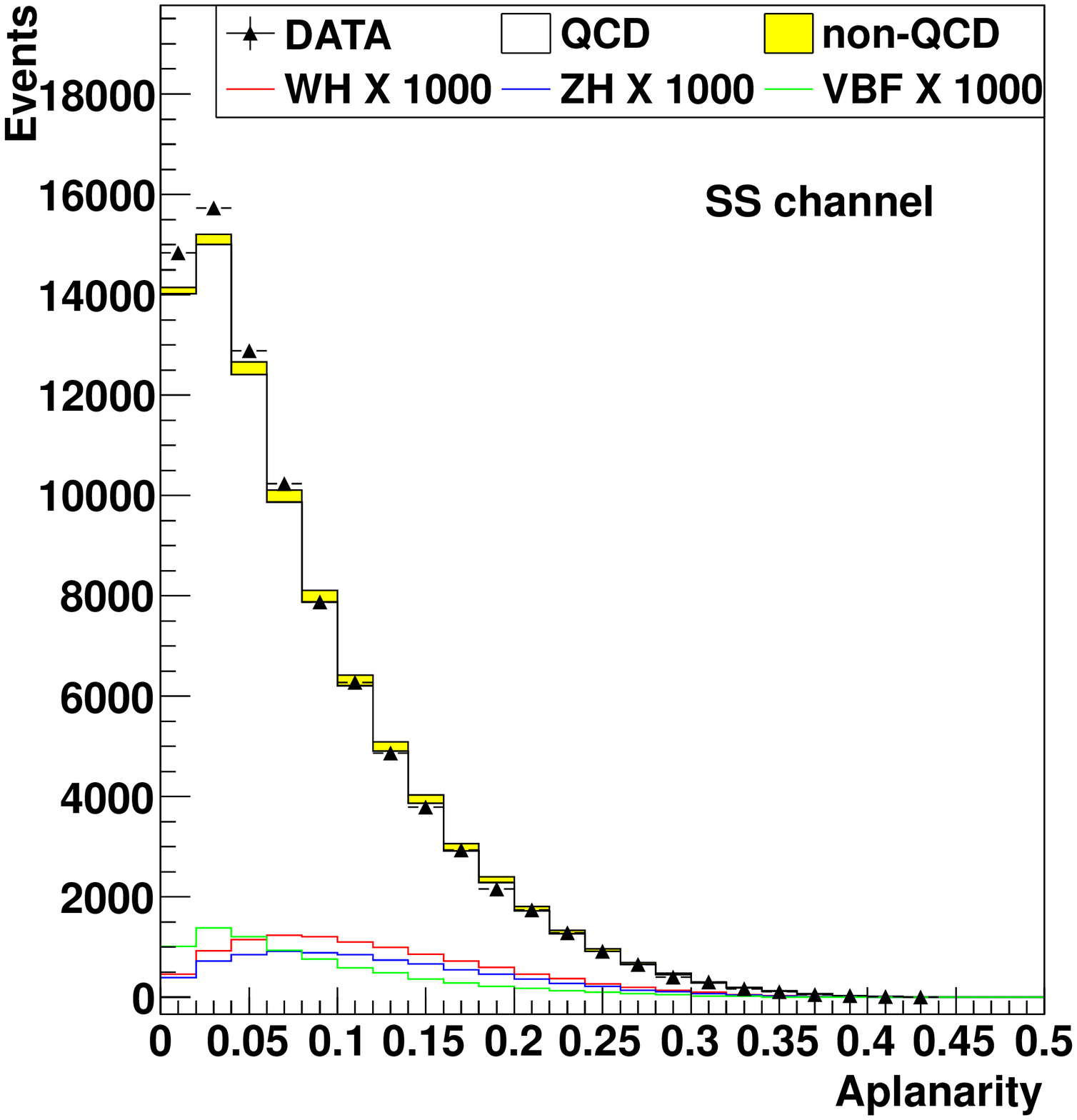}} \\
    \subfigure[]{\label{spher}\includegraphics[width=5.5cm]{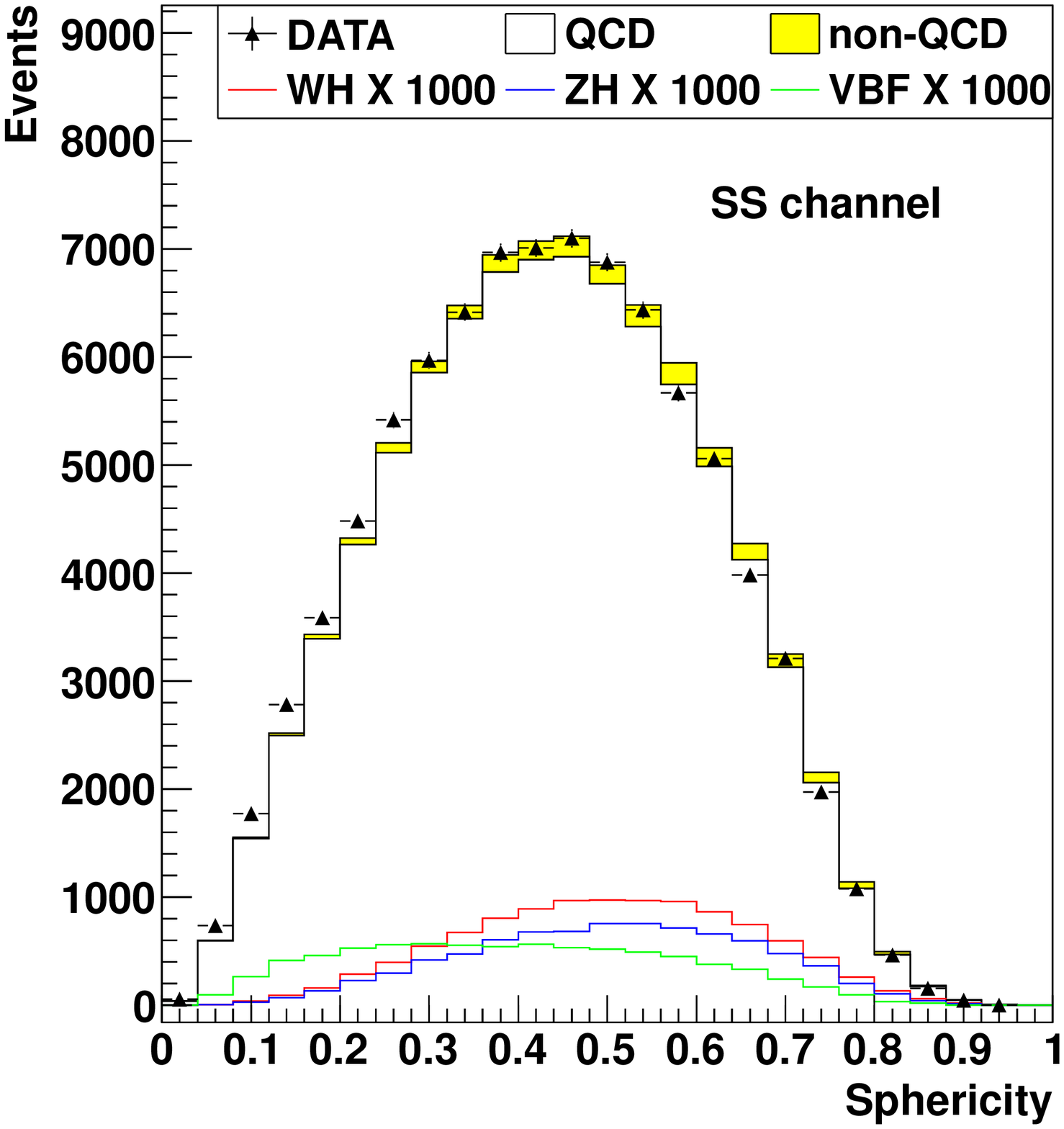}} 
    \subfigure[]{\label{centr}\includegraphics[width=5.5cm]{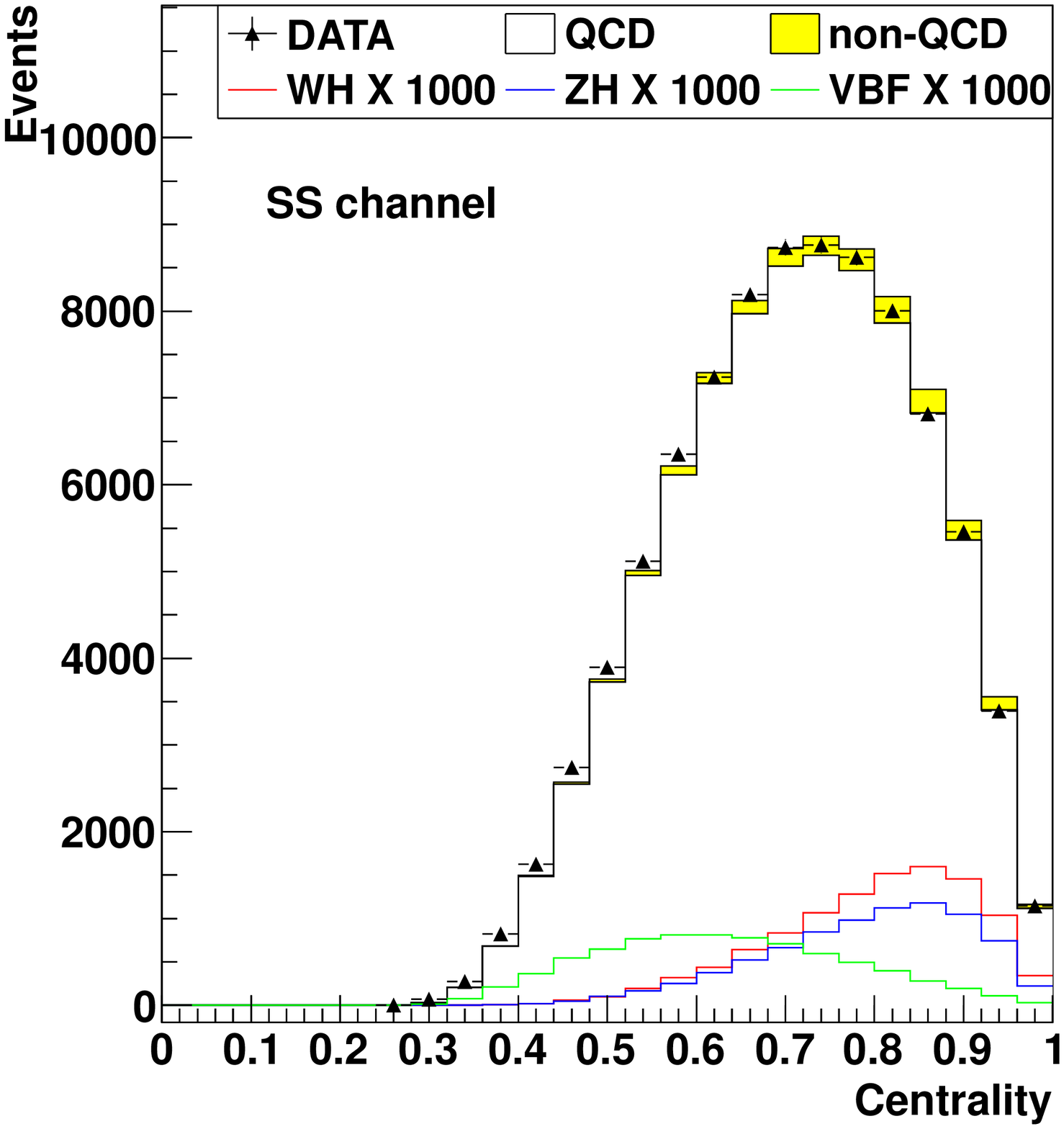}} \\
    \caption{The QCD multi-jet background prediction for the SS $b$-tag category of  
      \subref{eta_q1} the $\eta$ angle of the first leading untagged jet and          
      \subref{eta_q2} second leading untagged jet,     
      \subref{delta_eta_qq} $\Delta\eta$ of the two untagged jets,
    \subref{aplan} the aplanarity~\cite{Sjostrand:2000wi},
      \subref{spher} the sphericity~\cite{Sjostrand:2000wi}, and
      \subref{centr} centrality~\cite{Sjostrand:2000wi}.
       Descriptions of the signal and background histograms can be
       found in the caption of figure~\ref{FIG:TrainingVariables_SS_1}.      
      }
    \label{FIG:TrainingVariables_SS_3}
  \end{center}
\end{figure}

\begin{figure}[htbp]
  \begin{center}

    \subfigure[]{\label{deltaR_bb}\includegraphics[width=5.5cm]{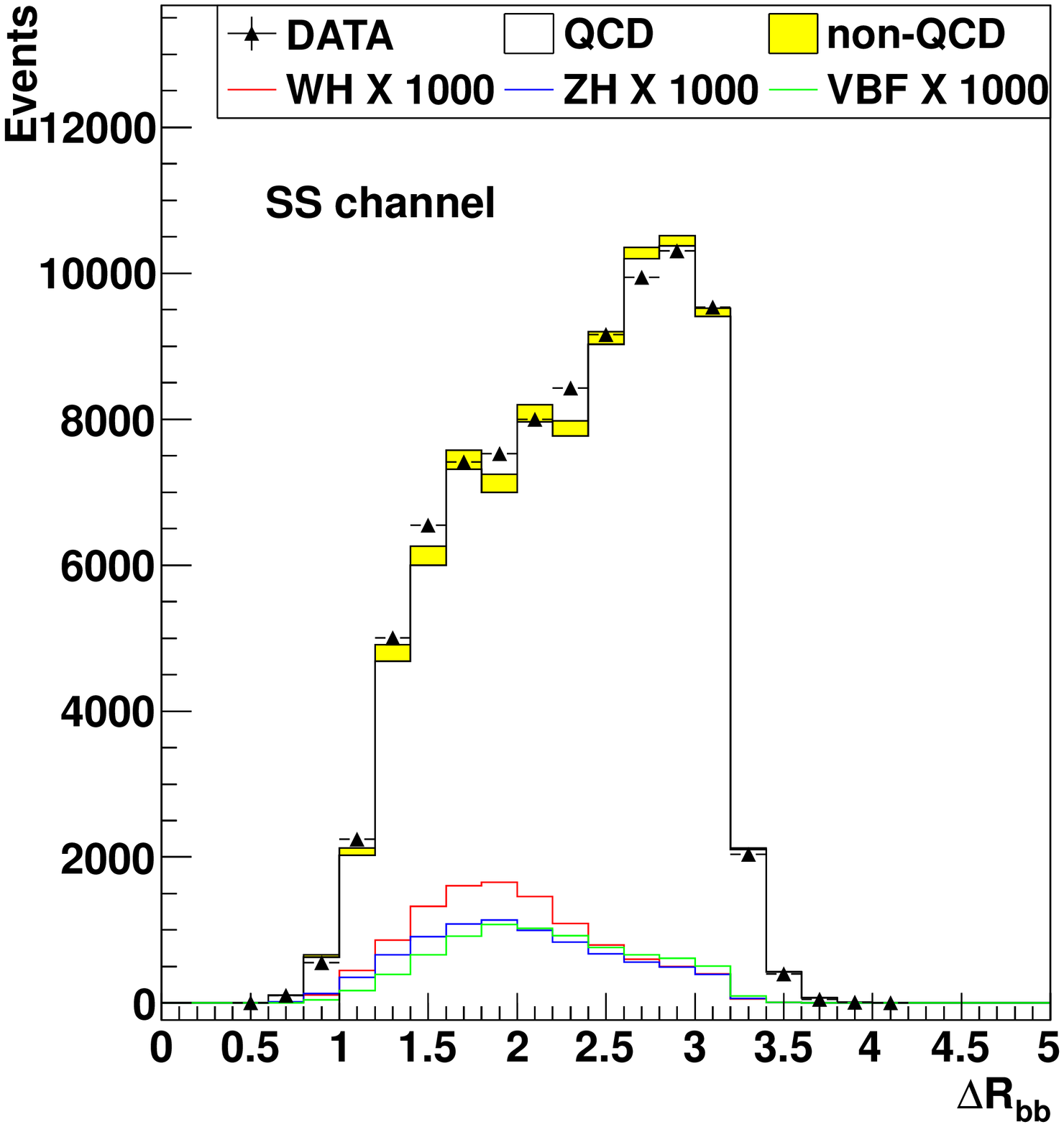}} 
    \subfigure[]{\label{deltaR_qq}\includegraphics[width=5.5cm]{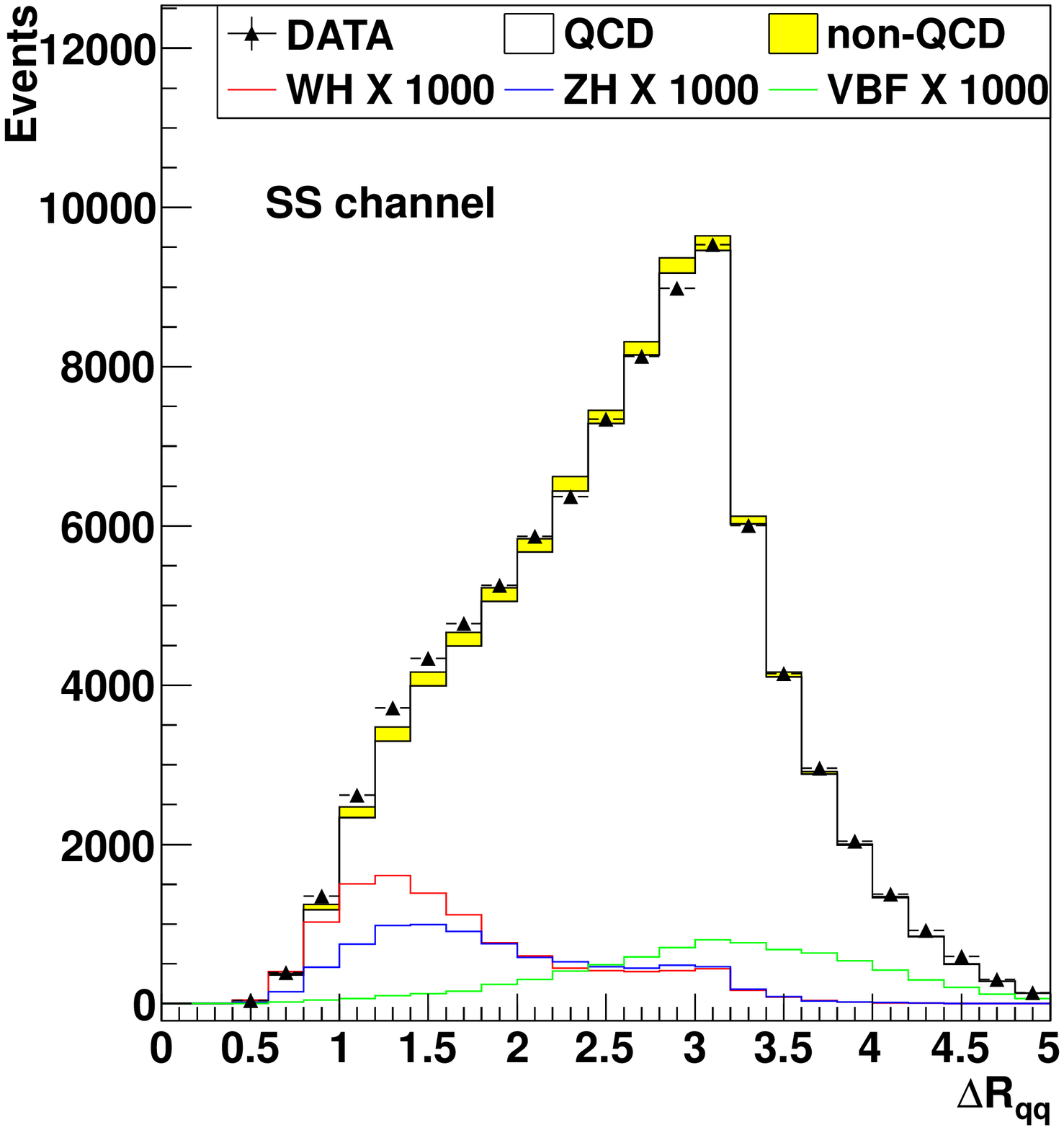}} \\
    \subfigure[]{\label{delta_phi_bb}\includegraphics[width=5.5cm]{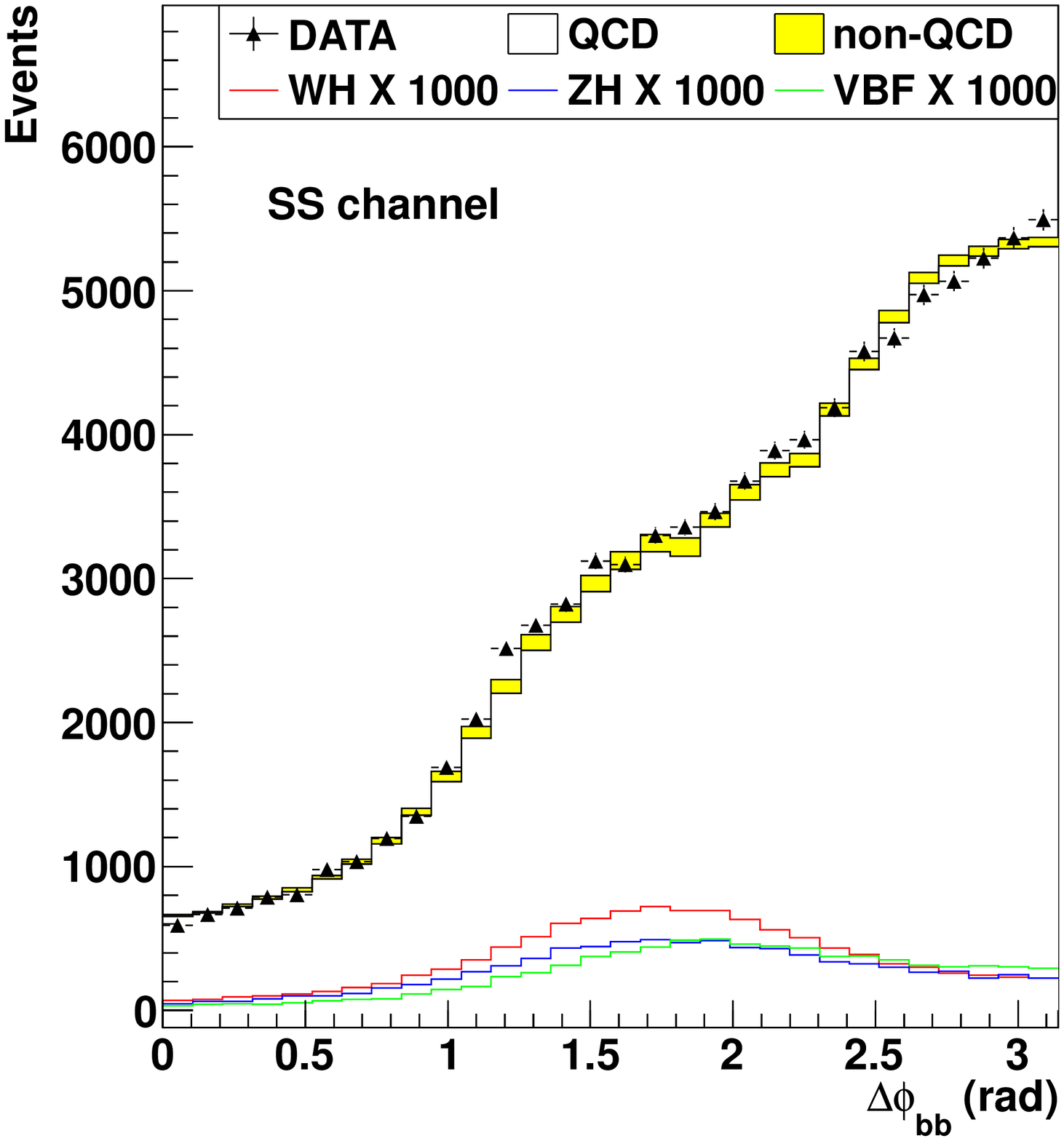}} 
    \subfigure[]{\label{delta_phi_qq}\includegraphics[width=5.5cm]{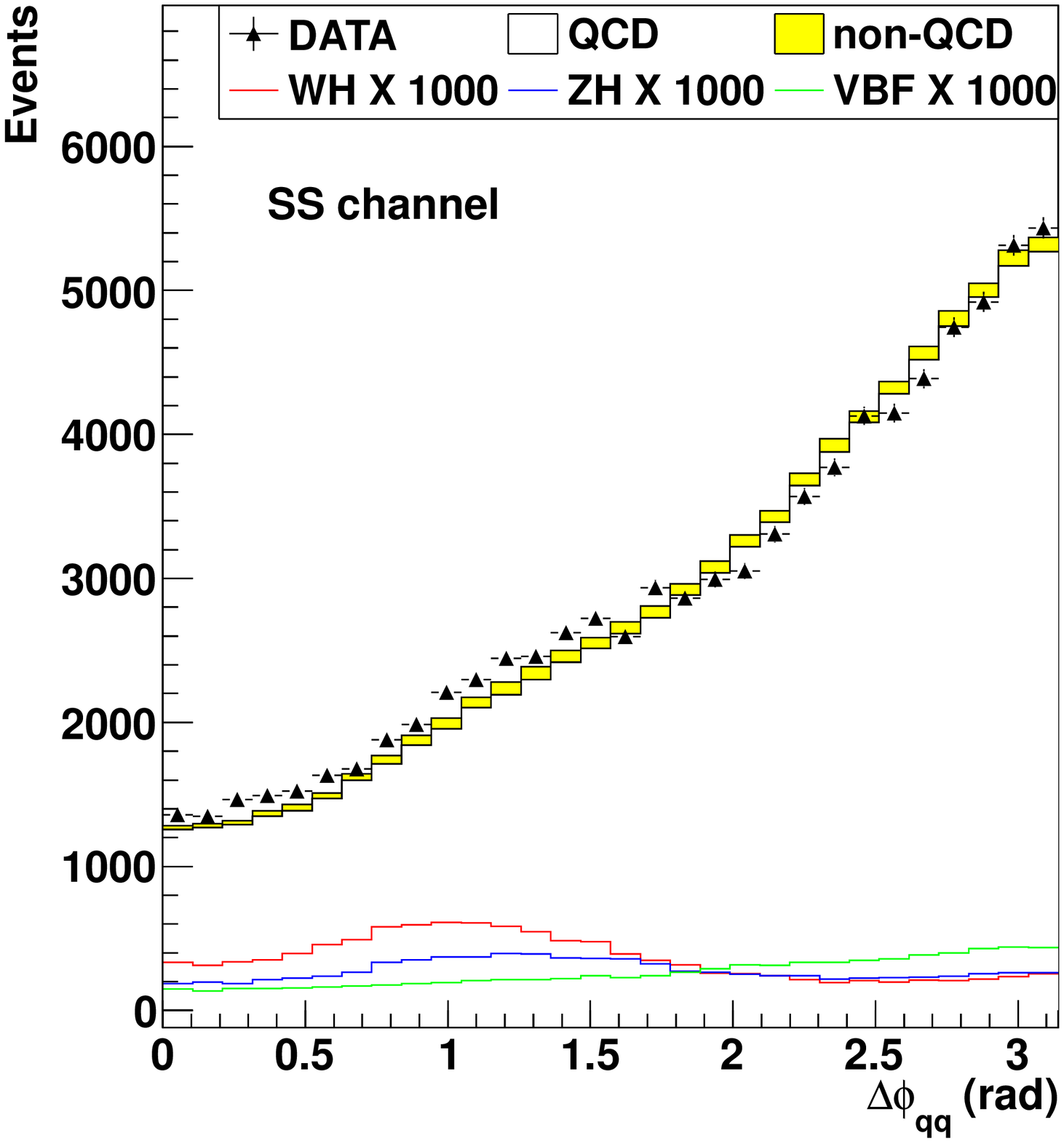}}\\
    \caption{The QCD multi-jet background prediction for the SS $b$-tag category of 
    \subref{deltaR_bb} the $\Delta R$ of the two $b$-tagged jets and \subref{deltaR_qq} of the two untagged jets, 
     \subref{delta_phi_bb} the $\Delta\phi$ of the two $b$-tagged jets  and
     \subref{delta_phi_qq} of the two untagged jets.
      Descriptions of the signal and background histograms can be
      found in the caption of figure~\ref{FIG:TrainingVariables_SS_1}.
     }
    \label{FIG:TrainingVariables_SS_4}
  \end{center}
\end{figure}

\begin{figure}[htbp]
  \begin{center}
    \subfigure[]{\label{YC_WH}\includegraphics[width=5.5cm]{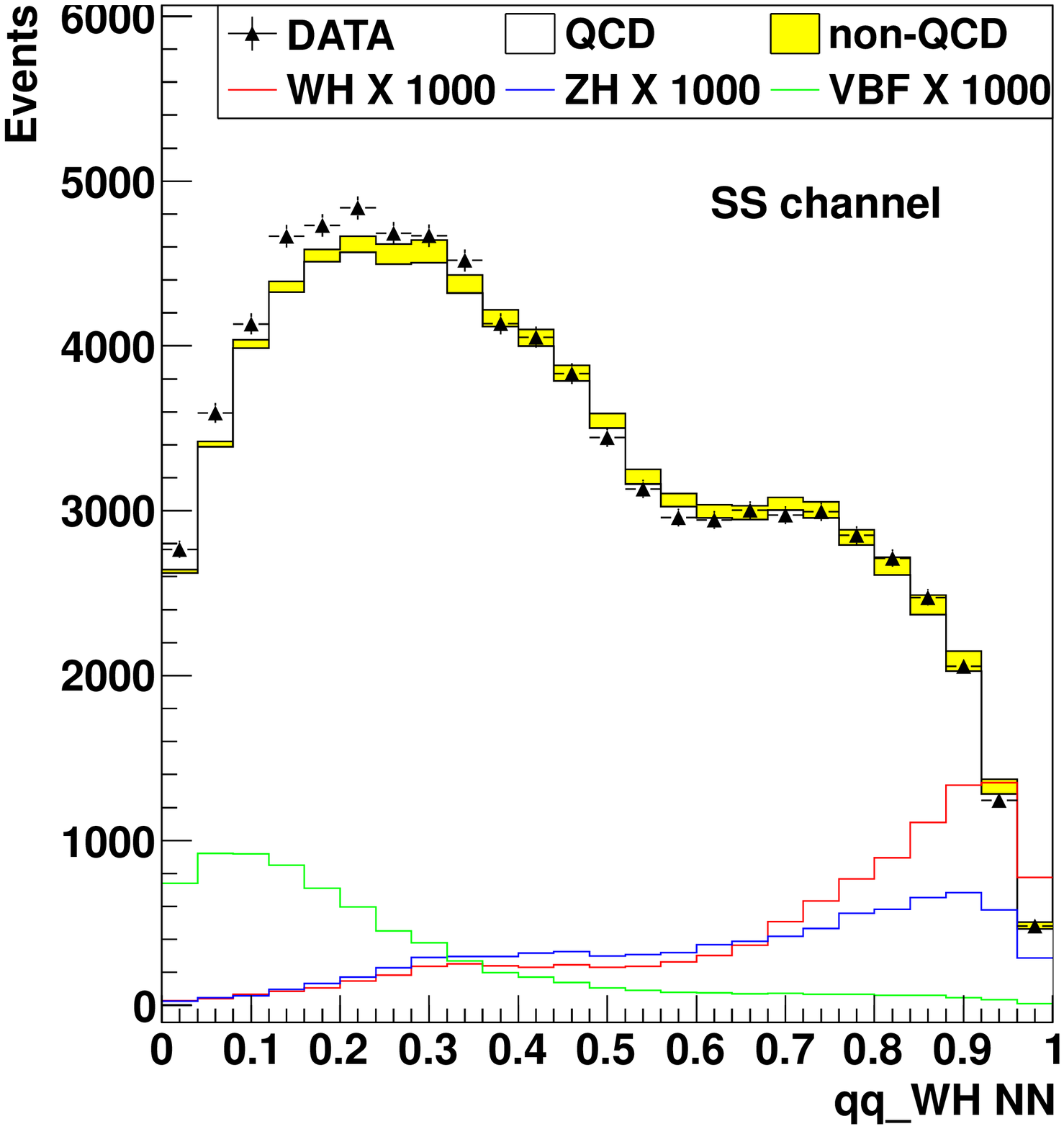}}
    \subfigure[]{\label{YC_ZH}\includegraphics[width=5.5cm]{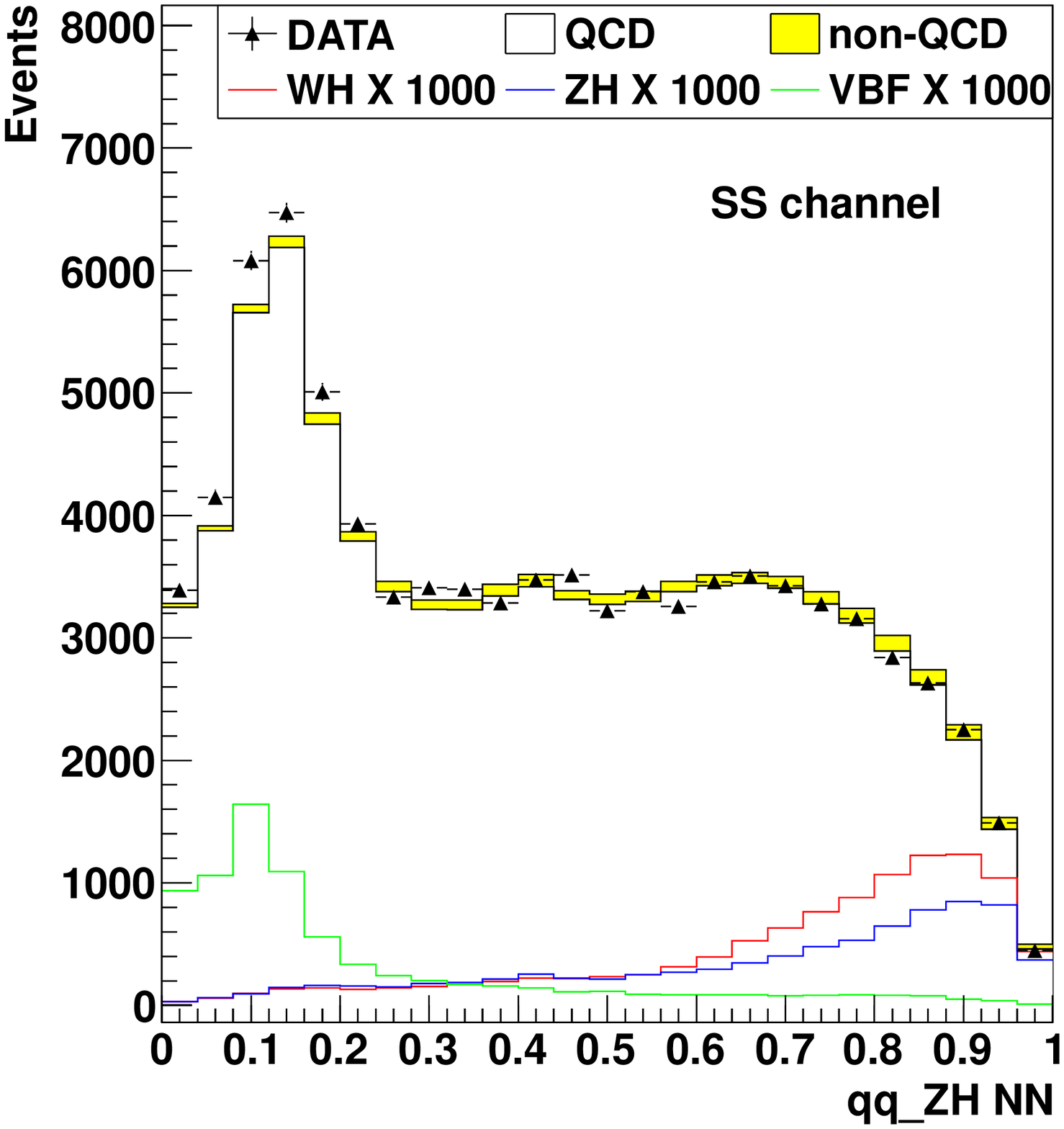}}
    \subfigure[]{\label{YC_VBF}\includegraphics[width=5.5cm]{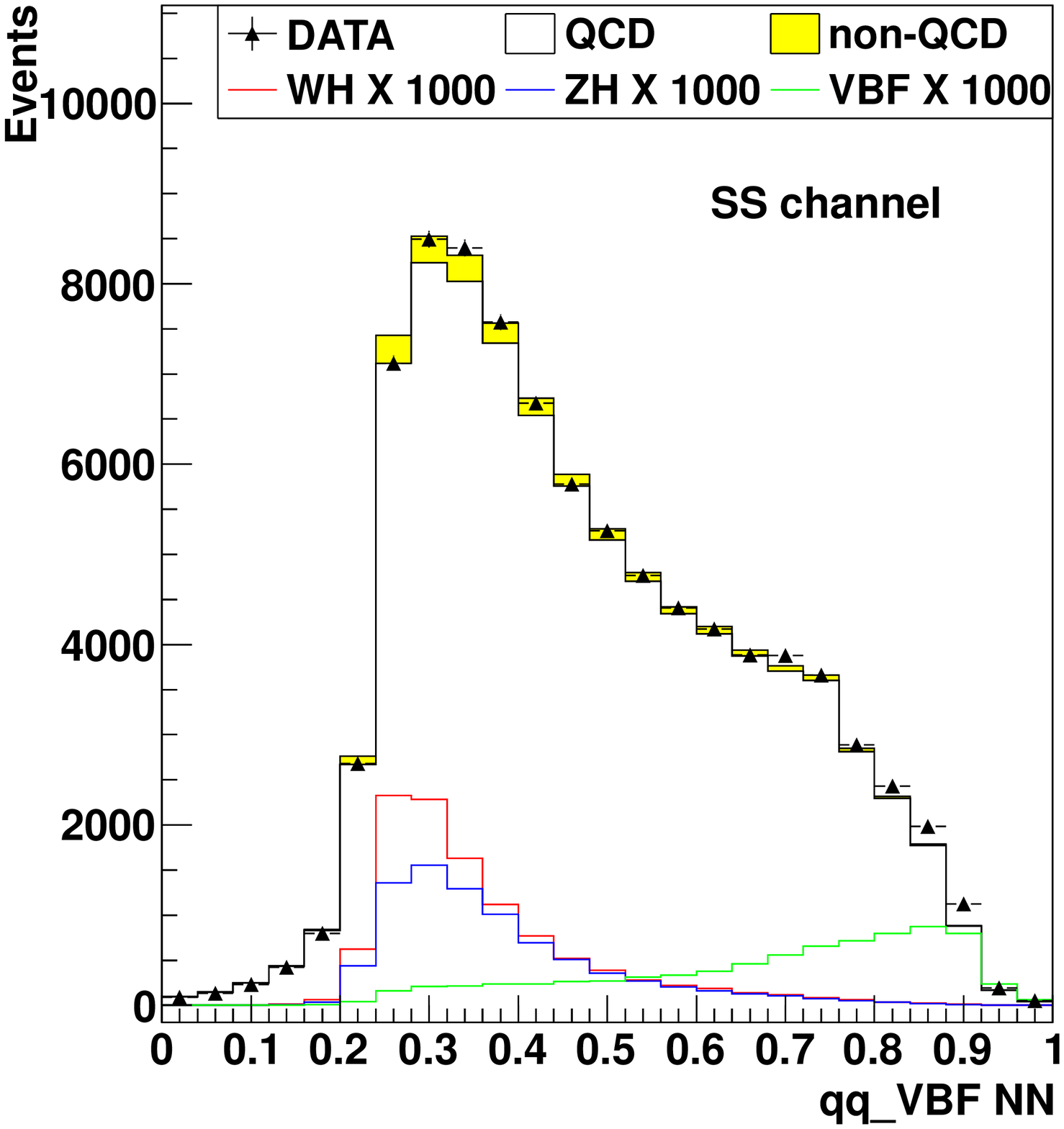}}
    \caption{The QCD multi-jet background prediction for the SS $b$-tag category of the 
     \subref{YC_WH} ~$qq\_WH$~NN, \subref{YC_ZH}~$qq\_ZH$~NN, and
      \subref{YC_VBF}~$qq$\_VBF~NN (section~\ref{SECTION:UntaggedJetsNeuralNetwork}).
       Descriptions of the signal and background histograms can be
       found in the caption of figure~\ref{FIG:TrainingVariables_SS_1}.      
      }
    \label{FIG:TrainingVariables_SS_5}
  \end{center}
\end{figure}

The distributions of the discriminating variables for the Higgs signal
and background are shown in figures~\ref{FIG:TrainingVariables_SS_1}-\ref{FIG:TrainingVariables_SS_5}.
The presence of a resonance in the {\it VH} and VBF channels, due to the
potential Higgs boson decay, produces a peak in the \mbb\, that is not
present in the QCD multi-jet background (figure~\ref{FIG:TrainingVariables_SS_1}\subref{mass_bb}).
A similar observation can be made for the \mqq\,
distribution in figure~\ref{FIG:TrainingVariables_SS_1}\subref{mass_qq}, where the {\it VH} channel features a resonance due to
the $W/Z$ decay (not observed in the VBF channel since the two $q$ jets
are not produced from the decay of a particle).
In
figures~\ref{FIG:TrainingVariables_SS_2}\subref{width_tower_q1}-\subref{width_tracks_q2}
the jet width distributions of untagged jets of the QCD multi-jet
background are broader than the Higgs signal due to the reason that is
described in section~\ref{SECTION:JetWidth}.
The two $q$ jets produced in the VBF events, which are produced from the two incoming quarks that each
radiates a weak boson,
tend to point in the forward but opposite directions. Thus the two $q$
jets are widely separated in the pseudo rapidity space.
These features are shown in
figures~\ref{FIG:TrainingVariables_SS_3}\subref{eta_q1}-\subref{delta_eta_qq}.
The {\it qq\_\!WH} NN (figure~\ref{FIG:TrainingVariables_SS_5}\subref{YC_WH}),
{\it qq\_ZH} NN
(figure~\ref{FIG:TrainingVariables_SS_5}\subref{YC_ZH}), and 
{\it qq\_}\!VBF NN
(figure~\ref{FIG:TrainingVariables_SS_5}\subref{YC_VBF}) distributions
are the outputs of three neural networks that are trained to separate the QCD multi-jet events
from WH, ZH and VBF productions, respectively.
%
%

Each variable demonstrates some ability to distinguish a Higgs boson from the background.   Some variables, such
as figures~\ref{FIG:TrainingVariables_SS_1}\subref{mass_4j} and  \ref{FIG:TrainingVariables_SS_5}\subref{YC_WH}
appear to have some mismodeling of the background. However the observed difference are 
within the shape uncertainties of the TRF QCD multi-jet prediction. 
These shape uncertainties are derived by testing these variables in the TAG, CONTROL, and NJET6 control regions
and propagating any difference as a shape uncertainty.


\clearpage

The {\it WH}-NN,  {\it ZH}-NN, and VBF-NN are trained using dedicated MC samples for signal modeling. 
A small subset (10\%) of single-tagged jet events, after random selection and application of the TRF, is 
used as the QCD multi-jet training sample.  The remaining 90\% of events are reserved for modeling
the NN output distributions.  As the shapes of the kinematic distributions are found to be consistent for 
both $b$-tagging categories, the NN is trained using SS events. 

The search focuses on Higgs boson mass hypotheses in the range $100 \leq m_H \leq 150\,\gevcc\,$
at  $5\,\gevcc\,$ intervals. The sensitivity of the search is improved by using separate trainings at three specific
Higgs boson masses: 100\,\gevcc, 120\,\gevcc, and 140\,\gevcc.  For each Higgs boson mass hypothesis, we choose the training that gives the best search sensitivity.

\begin{figure}
  \centering
  \subfigure{\includegraphics[width=10cm]{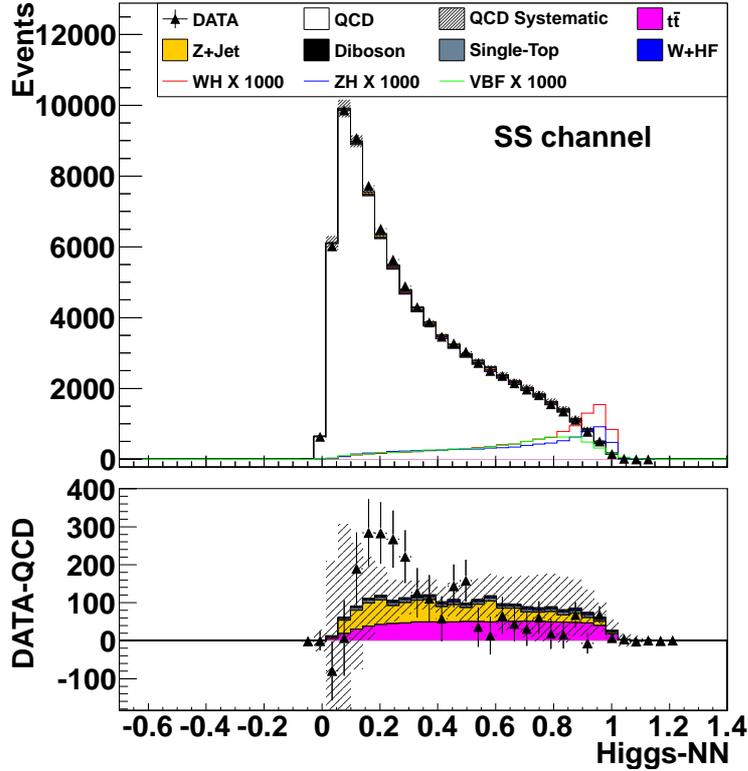}}
  \hspace{1cm}
  \caption{\SD\, distribution of 125\,\gevcc\, Higgs boson events with
    both $b$ jets tagged by SecVtx, after the VBF-NN correction function was applied.
    All backgrounds are stacked and the superimposed Higgs boson signal is scaled
    by 1000.  As the QCD multi-jet background is large, plots of the difference
     of data and QCD multi-jet are plotted with a stacked plot of non-QCD 
    background and QCD multi-jet systematic uncertainty. Both plots show the data are consistent
    with the background, especially at large Higgs-NN score where the higgs signal peaks.
    }
  \label{FIG:MH125NN}
\end{figure}

Only variables that are well modeled by the TRF are used to train the {\it WH}-NN, {\it ZH}-NN, and 
VBF-NN.  
As a further validation, the modeled outputs of the {\it WH}, {\it ZH}, and VBF networks are compared to TAG events in data.
The {\it WH}, {\it ZH} networks are found 
to be well modeled, but the VBF-NN requires an additional correction, analogous to the  re-weighting 
performed to correct \mqq\, (section~\ref{SECTION:QCDMODEL}). 
Figure~\ref{FIG:MH125NN} shows the \SD\,
distribution of 125\,\gevcc\, Higgs boson events with both $b$ jets
tagged by SecVtx,
after the VBF-NN correction function was applied. The histogram shows
the data, a stacked distribution of the backgrounds, and the Higgs boson signal
scaled by 1000.
As the QCD multi-jet background is large, it is difficult to see the non-QCD contributions and the QCD  
uncertainty.  
In the lower QCD subtracted data plot, it is easier to see how well the background is modeled. 
%
%
This plot shows the QCD uncertainty is as large as the total non-QCD 
contributions and the QCD subtracted data is consistent with the non-QCD background and the QCD uncertainty. 



%




\section{Systematic uncertainties}~\label{SEC:Systematics}
This search considered systematic effects that affect the
normalization (rate systematic uncertainty) and the output (shape
systematic uncertainty) of the \SD\,for the signal and
background. The rate systematic uncertainties 
are defined as the variations of the number of events that pass the 
selection requirements. The shape-related systematic uncertainties are expressed
as fractional changes in the binned distributions.

The systematic effects that affect the normalization of the
Higgs boson and non-QCD background are the uncertainty on the jet
energy scale (JES)~\cite{Bhatti:2005ai}, on the PDF, $b$-tagging scale factor, initial and final
state radiation (ISR and FSR), trigger efficiency, integrated luminosity,
and cross section~\cite{Tevatron:Higgs}.
The effects that affect the shape of the Higgs boson and non-QCD backgrounds 
are the uncertainties on the JES, ISR, FSR, and the jet width.
%
The shape uncertainties are evaluated by adjusting 
their values by $\pm 1\sigma$, 
and  
propagating this change through the event selection and \SD.
Table~\ref{TABLE:SystematicsSummary} summarizes all systematic uncertainties.

%
%

Only shape uncertainties are considered for the QCD multi-jet component, the normalization is unconstrained. 
The TRF QCD shape uncertainties arise from uncertainties in the interpolation, \mqq\, and VBF-NN correction functions. 
The TRF shape uncertainty is defined as the shape difference of the
nominal QCD shape and a systematically shifted version.



The interpolation uncertainty accounts for sample-dependence of the
TRF. A TRF is measured in the TAG region to its application in the signal region.
%
%
Another TRF is measured in the \Ctrl\, region
(figure~\ref{FIG:MbbMqqRegion}) and is applied to the signal region.  
The shape difference of the nominal \Tag\, TRF and the \Ctrl\, TRF defines the interpolation uncertainty. 

The \mqq\, and VBF-NN distributions require an additional correction to improve their TRF modeling (sections~\ref{SECTION:QCDMODEL} and \ref{SECTION:NNTraining}). The nominal correction functions are measured in
the \Tag\, region and an alternative is measured in the \Ctrl\, (\mqq)  and
\Njet\, (VBF-NN) regions. 
The shape difference between the usage of the nominal and alternative correction function defines the correction function shape uncertainty.
%
%
%
%


\begin{table}[h!]
\begin{center}
\begin{tabular}{|l|l|}  \hline
  \multicolumn{2}{|l|}{ TRF (QCD multi-jet) uncertainties} \\ \hline
  TRF interpolation              & Shape  \\ 
  TRF \mqq\, correction                 & Shape  \\ 
  TRF VBF-NN correction     & Shape \\  \hline \hline
  \multicolumn{2}{|l|}{ Signal and Background uncertainties} \\ \hline
  Luminosity                    & $\pm$ 6\% Rate \\ 
  Trigger                          & $\pm$ 3.55\%  Rate \\  
  SecVtx+SecVtx            & $\pm$ 7.1\% Rate \\ 
  SecVtx+JetProb          & $\pm$ 6.4\% Rate \\ 
  Jet Energy Correction & $\pm$ 9\% Rate \\ 
  & Shape        \\ 
  Jet width                  & Shape \\ \hline \hline
  \multicolumn{2}{|l|}{Cross section uncertainties} \\  \hline
  \ttbar\, and single-top & $\pm$ 7\% Rate \\ 
  Diboson (WW/WZ/ZZ)  & $\pm$ 6\% Rate \\
  {\it W}+$\mathit{HF}$ and {\it Z}+jets          & $\pm$ 50\% Rate \\
  {\it WH}/{\it ZH}                      & $\pm$ 5\% Rate  \\
  VBF                      & $\pm$ 10\% Rate  \\ \hline \hline
  \multicolumn{2}{|l|}{Signal uncertainties} \\ \hline
  PDF                                & $\pm$ 2\% Rate \\ 
  ISR/FSR                       & $\pm$ 3\% Rate \\
  & Shape \\ \hline
\end{tabular}
\end{center}
\caption{Summary of all systematic uncertainties.}
\label{TABLE:SystematicsSummary}
\end{table}

\section{Results}
\begin{figure}
  \centering
  \includegraphics[height=10cm]{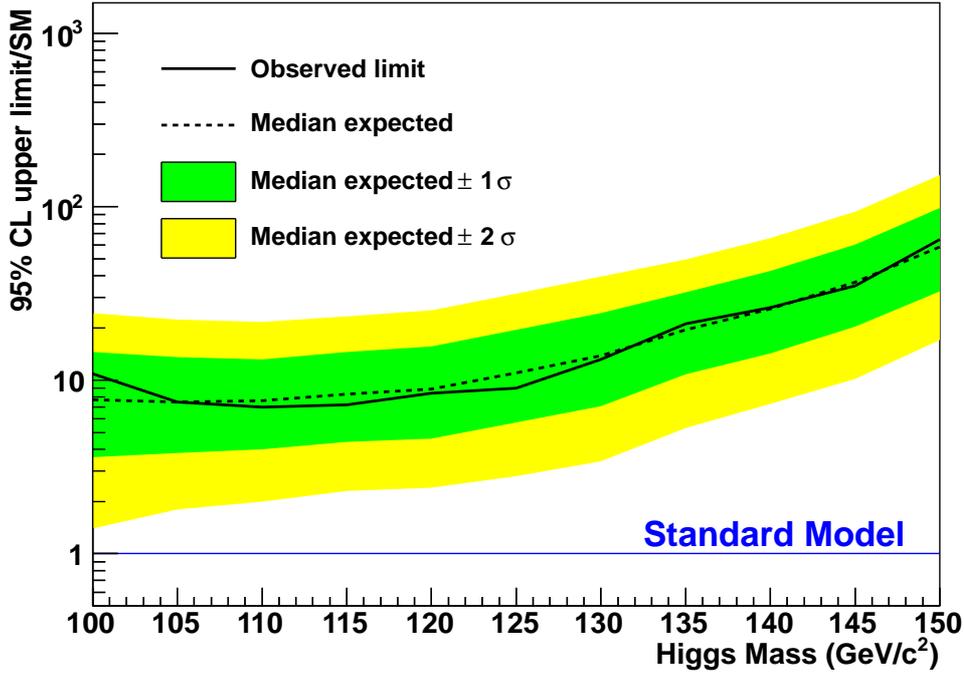}
  \caption{Upper limits at 95\% CL for combined SS and SJ channels:
    the expected and observed limits are plotted as a function of the
    Higgs boson mass. The limits are relative to the expected SM Higgs boson production, which includes the $H \rightarrow
    \bbbar$ branching ratio.}
  \label{FIG:CombinationLimit}
\end{figure}

The Higgs-NN output distribution in data is compared to the background predictions.
No evidence of a Higgs boson signal is found, nor any 
disagreement
between the predicted background and observed data. Upper exclusion limits are calculated on 
the Higgs boson cross-section at the 95\% CL.  
The limits are calculated using a Bayesian method with a non-negative flat
prior for the signal cross section. We integrate over Gaussian priors
for the systematic uncertainties, truncated to ensure that no prediction is negative,
 and incorporate correlated rate and
shape uncertainties as well as uncorrelated bin-by-bin statistical
uncertainties~\cite{PhysRevLett.104.061802}.
Figure~\ref{FIG:CombinationLimit} and
table~\ref{TABLE:ALL_Limits} show the 
limits from the combination of SS and SJ $b$-tagging categories.
The observed limits agree with the expected limits.

\begin{table}
  \begin{center}
    \begin{tabular}{|c|l|l|l|l|l|l|}
      \hline
      Higgs mass (\gevcc) & $-2\sigma$ & $-1\sigma$ & Median & $+1\sigma$ & $+2\sigma$ & Observed\\
      \hline
      100 & 1.4    &  3.6    & 7.7     & 14.5    & 24.4    &  10.9  \\
      105 & 1.8    &  3.8    & 7.5     & 13.6    & 22.3    &  7.5    \\
      110 & 2.0    &  4.0    & 7.6     & 13.2    & 21.7    &  7.0    \\
      115 & 2.3    &  4.4    & 8.3     & 14.5    & 23.4    &  7.2    \\
      120 & 2.4    &  4.6    & 8.9     & 15.6    & 25.3    &   8.4   \\
      125 & 2.8    &  5.7    & 11.0   & 19.5    & 31.6    &   9.0   \\
      130 & 3.4    &  7.1    & 13.8   & 24.3    & 39.5    &  13.2  \\
      135 & 5.3    & 10.8   & 19.5   & 32.2    & 49.6    &  21.2  \\
      140 & 7.3    & 14.3   & 25.8   & 42.7    & 66.1    &  26.2  \\
      145 & 10.2  & 20.4   & 36.7   & 60.5    & 93.4    &  35.1  \\
      150 & 17.1  & 32.5   & 58.7   & 98.2    & 152.0  &  64.6  \\\hline
    \end{tabular}
  \end{center}
  \caption{Expected and observed 95\% CL upper limits for the combined SS and SJ
    channels. The limits are relative to the expected Higgs cross section.}
  \label{TABLE:ALL_Limits}
\end{table}



%
\section{Summary}
A search for the Higgs boson is performed in the all-hadronic final
state using 9.45\invfb\, of data collected by the CDF II detector. 
The results discussed in this paper have halved the expected limit of
the previous search~\cite{PhysRevD.84.052010}.  
Half of the improvement comes from additional data and the expanded
signal region contributes an additional 17\%.
The reduction of the $b$-jet energy resolution by 18\%, adding a new
jet width measurement, improving the QCD multi-jet modeling,  and
adding more variables in the Higgs neural network and improving its
training contributes another 24\%.
The combination of multi-variate techniques improved the best signal-to-background ratio
from 0.0007, if the \mbb\, distribution alone was used for the search, to 0.006, which is almost a ten-fold increase. 
No significant Higgs boson signal is observed and upper exclusion
limits are set on the observed Higgs cross section relative to the SM
rate as a function of Higgs boson mass in the range 100-150\,\gevcc.
For a 125\,\gevcc\, Higgs boson, the 95\% CL expected (observed) limit is 11.0 (9.0)
times the expected SM rate.
This search is CDF's fourth most sensitive $H \rightarrow \bbbar$ search and is more sensitive than
CDF's $\ttbar H$~ \cite{Collaboration:2012bk} and similar to 
CDF's $H \rightarrow \gamma\gamma$~\cite{Aaltonen2012173} searches, which have an expected limit of 12.6 and  
9.9 for a 125\,\gevcc\, Higgs boson, respectively.  
%
%
CDF has also developed an improved algorithm to identify (tag)  $b$-jets~\cite{Freeman201364},  
which improves the $b$-tagging rate from 39\% to 54\% and 
was used in the latest $ZH \rightarrow \ell\ell \bbbar$~\cite{PhysRevLett.109.111803} and 
$WH\rightarrow \ell \nu \bbbar$~\cite{PhysRevLett.109.111804} searches.
The addition of new $b$-jet tagger could potentially improve this search's expected limit by 
an additional 40\% which would lower the expected limit to 7.9 times the expected SM rate for a 125\,\gevcc\, Higgs
boson.   
The all-hadronic search is a unique channel at the Tevatron that has not been explored at the LHC.  
The improvements described in this paper, such as the data-driven QCD multi-jet prediction, 
$b$-jet energy corrections, jet width, and two-stage NN can be applied to $H\rightarrow \bbbar$ searches and
other multi-jet analyses at the LHC.
\acknowledgments{
We thank the Fermilab staff and the technical staffs
of the participating institutions for their vital contributions.
This work was supported by the U.S. Department
of Energy and National Science Foundation; the
Italian Istituto Nazionale di Fisica Nucleare; the Ministry
of Education, Culture, Sports, Science and Technology
of Japan; the Natural Sciences and Engineering
Research Council of Canada; the National Science Council
of the Republic of China; the Swiss National Science
Foundation; the A.P. Sloan Foundation; the Bundesministerium
f\"{u}r Bildung und Forschung, Germany; the Korean
World Class University Program, the National Research
Foundation of Korea; the Science and Technology
Facilities Council and the Royal Society, UK; 
the Russian Foundation for Basic Research;
the Ministerio de Ciencia e Innovaci\'{o}n, and Programa
Consolider-Ingenio 2010, Spain; the Slovak R\&D Agency;
the Academy of Finland; and the Australian Research Council (ARC).
}

\bibliography{Reference}

\end{document}